\documentclass[12pt,prd]{revtex4}

\usepackage{amsmath}
\usepackage{graphicx,color}

\newcommand{\arxiv}[1]{{arXiv:#1}}

\newcommand{\tr}{\rm{tr}\,}

\newcommand{\PpT}{(P^+_\mu)^\mathrm{T}}
\newcommand{\PmT}{(P^-_\mu)^\mathrm{T}}
\newcommand{\T}{\mathrm{T}}
\newcommand{\mrm}{\mathrm}
\newcommand{\iu}{\mathrm{i}}

\newcommand{\A}{X}
\newcommand{\B}{Y}

\begin{document}

\title{
Phase structure for lattice fermions
\\ with flavored chemical potential terms
}

\author{Tatsuhiro Misumi}
\email{tmisumi@bnl.gov}
\affiliation{Physics Department, Brookhaven National Laboratory, 
Upton, NY 11973, USA}

\begin{abstract}
We discuss the chiral phase diagram in the parameter space 
of lattice QCD with minimal-doubling fermions,
which can be seen as lattice fermions with flavored chemical potential terms.
We study strong-coupling lattice QCD with the Karsten-Wilczek formulation, 
which has one relevant parameter $\mu_{3}$ as well as gauge coupling
and a mass parameter. 
We find a nontrivial chiral phase structure with a second-order phase transition 
between chiral symmetric and broken phases. 
To capture the whole structure of the phase diagram, we study the
related lattice Gross-Neveu model. The result indicates that the chiral phase 
transition also exists in the weak-coupling region.
From these results we speculate on the $\mu_{3}$-$g^{2}$ chiral phase diagram 
in lattice QCD with minimal-doubling fermions, and discuss their application
to numerical simulations. 
\end{abstract}

\maketitle

\newpage


\section{Introduction}
\label{sec:Intro}

The famous no-go theorem \cite{KarS,NN} states that lattice 
fermion actions with chiral symmetry, locality and other common 
features must produce massless degrees of freedom in 
multiples of two in a continuum limit.  This is contradictory with a 
phenomenological fact that there exist only three quarks with masses 
below the QCD scale.  By now several fermion constructions to bypass the no-go theorem
have been developed, although all of them have their individual
shortcomings: The explicit chiral symmetry breaking with
the Wilson fermion approach \cite{Wil} results in an additive mass
renormalization, which in turn requires a fine-tuning of the mass
parameter for QCD simulations. Domain-wall \cite{Kap, FuSh}
and overlap fermions \cite{GW, Neu} produce a single fermion mode 
by modifying the definition of chiral symmetry or introducing 
the momentum-dependent chiral charge, although they lead to 
rather expensive simulations algorithms..  
These approaches attempt to realize single fermionic degrees of freedom 
by breaking the requisite conditions for the no-go theorem.
On the other hand, there is another direction to approach numerical
simulations.  According to \cite{KarS}, Hypercubic symmetry and
reflection positivity of actions result in $2^{d}$ species of fermions
where $d$ stands for the dimension.  Thus it is potentially possible
to reduce the number of species by breaking hypercubic symmetry
properly.  Actually, the staggered fermion approach \cite{KS, Suss,
KaS,  Sha, GS}, with only 4 species of fermions does this and possesses
flavored-hypercubic symmetry instead.  However this requires rooting
procedures for the physical $2$ or $(2+1)$-flavor QCD simulation.

A possible goal in this direction is a lattice fermion with $2$ species, the
minimal number required by the no-go theorem.  Such a minimal-doubling 
action was first proposed by Karsten, and later by Wilczek \cite{Kar, Wilc}. 
Other than the original type, two more types are known as Creutz-Borici 
type \cite{Creutz1, Bori, Creutz2} and Twisted-ordering type \cite{CM}.  
These fermions all possess one exact chiral symmetry and ultra locality.  
As such they could be faster for simulation, at least 
for two-flavor QCD, than other chirally symmetric lattice fermions.
However it has been shown \cite{Bed1, Bed2, KM1, KM2} that we need to
fine-tune several parameters for a continuum limit with these actions.
This is because they lack sufficient discrete symmetry to prohibit
redundant operators from being generated through loop
corrections \cite{Cichy, Cap1, Cap2, Cap3}.  Thus the minimally doubled
fermions have not been extensively used so far.  Nevertheless, there
is the possibility to apply them to simulations if one can efficiently
perform the necessary fine-tuning of parameters.

In this paper we pursue the chiral phase structure in the parameter plane 
for minimal-doubling lattice QCD for two purposes:
One purpose is to understand properties of these formulations theoretically. 
The other purpose is to show their applicability to 
lattice QCD simulations since understanding a nontrivial phase diagram, 
as with the Aoki phase in Wilson fermion \cite{AokiP, AokiU1, SS, Creutz3, ACV}, 
can be useful in showing the applicability to lattice QCD.
We first show that minimal-doubling fermions can be seen as a special case of
lattice fermions with species-dependent (imaginary) chemical potential terms. 
We focus on the Karsten-Wilczek formulation with one relevant parameter $\mu_{3}$, 
which corresponds to a mass parameter in the analogy of Wilson fermion.
We next investigate the chiral phase structure in the space of the gauge coupling 
and this relevant parameter. 
We analyze strong-coupling lattice QCD, and show that chiral symmetry
is spontaneously broken in a certain range of the parameter while
the chiral condensate is zero outside the range.
We find a second-order phase transition between these chiral symmetric 
and broken phases. We also show that pion
becomes massless as a Nambu-Goldstone boson in the chiral-broken phase
while the sigma meson becomes massless on the second-order phase boundary
due to the critical behavior of the second-order phase transition.
We investigate the lattice Gross-Neveu model to capture an entire phase structure.
From these results we suppose a similar chiral phase structure in 4d QCD
with flavored-chemical-potential lattice fermions, and discuss their applicability
to lattice QCD simulations.

In Sec.~\ref{sec:FCP}, we study lattice fermions with flavored chemical
potential, or minimal-doubling fermions.
In Sec.~\ref{sec:SLQ}, we investigate a chiral phase structure 
in the framework of strong-coupling lattice QCD.
In Sec.~\ref{sec:GN}, we study the Gross-Neveu model to obtain
information of the whole phase diagram.
In Sec.~\ref{sec:QCD}, we discuss a phase structure in 4d QCD 
from the results of the last two sections.   
Section \ref{sec:SD} is devoted
to a summary and discussion.


\section{Flavored-chemical-potential fermion}
\label{sec:FCP}

In this section we study minimal-doubling fermions
and their generalization.
Before going to the main theme, let us remind ourselves
of Wilson's prescription to shirk the no-go theorem.
Wilson fermion extracts one light fermion by introducing
a species-dependent mass term, which we call a ``flavored-mass term".
A free action and a Dirac operator of Wilson fermion are given by
\begin{equation}
S_{\rm W}\,=\,a^{4}\sum_{n,\mu}\bar{\psi}_{n}\gamma_{\mu}{\psi_{n+\mu}-\psi_{n-\mu}\over{2a}}+
a^{5}\sum_{n,\mu}\bar{\psi}_{n}{2\psi_{n}-\psi_{n+\mu}-\psi_{n-\mu}\over{2a^2}},
\label{WS}
\end{equation}
\begin{equation}
aD_{\rm W}(p)=i\gamma_{\mu}\sin p_{\mu}a\,+\,\sum_{\mu}(1-\cos p_{\mu} a).
\label{WD}
\end{equation}
We here exhibits a lattice spacing $a$ to manifest mass dimensions
of each term.
In this formulation, 15 out of 16 species have $O(1/a)$ mass and are decoupled
in the naive continuum limit.
As shown in \cite{CKM1}, this is not the only case of flavored-mass terms.
There are four types of nontrivial flavored-mass terms
($M_{\rm f}=M_{\rm P}$, $M_{\rm V}$, $M_{\rm A}$, $M_{\rm T}$), 
which satisfy $\gamma_{5}$ hermiticity and give second-derivative terms up to 
$O(a^2)$ errors as with the usual Wilson term.
A general form of Wilson-type fermions are written as
\begin{equation}
aD_{\rm fm}(p)=i\gamma_{\mu}\sin p_{\mu}a\,+M_{\rm f}(p), 
\label{FM}
\end{equation}
where $M_{\rm f}(p)$ stands for flavored-mass terms.
Details of species-splitting depends on explicit forms of $M_{\rm f}(p)$ \cite{CKM1}.

\begin{figure}
\includegraphics[height=7cm]{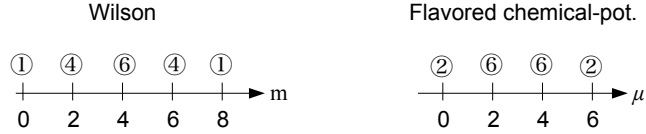} 
\caption{Species-splitting in Wilson and Karsten-Wilczek fermion.
Circled numbers stand for the number of massless flavors on each point.}
\label{MD}
\end{figure}

We now consider further deformation of the fermion action.
We multiply $M_{\rm f}$ by $\gamma_{4}$ or $i\gamma_{4}$ as
\begin{equation}
aD_{\rm fc}(p)=i\gamma_{\mu}\sin p_{\mu}a\,+(i)\gamma_{4}M_{\rm f}(p).
\label{FCP}
\end{equation}
In these cases, degeneracy of species is lifted by specie-dependent real or imaginary
chemical potential terms, not by species-dependent mass.
It means that we can get rid of some doublers by this method too.
We name such terms as ``flavored-chemical-potential terms",
and name lattice fermions with them as ``flavored-chemical-potential (FCP) fermions".
(We will later discuss problems with this kind of naive introduction of chemical potential \cite{HK}.)
It is obvious that a real type of FCP terms
breaks down $\gamma_{5}$ hermiticity and leads to a sign problem
while an imaginary type keeps it and has no sign problem.
An outstanding point in this formulation is that ultra-locality and one exact $U(1)$ 
chiral symmetry remains intact as
\begin{equation}
\{\gamma_{5},D_{\rm fc}(p)\}=0.
\end{equation}
(See Refs.~\cite{CKM1, T, Saidi, Saidi2, CSR} for details of chiral symmetry in this type of lattice fermions.)
In principle, by using this deformation we can reduce 16 species
to smaller multiple numbers of two without losing all chiral symmetries.
We note that the chemical potential term, of course, breaks hypercubic symmetry 
into cubic symmetry and breaks C, P, T into CT and P \cite{Bed1,Bed2, KM1,KM2} 
as far as $M_{f}(p)$ is cubic-symmetric. 
It means that this formulation automatically corresponds to 
finite-density systems unless we tune several parameters.
In this paper we concentrate on the following explicit form of imaginary-type FCP fermions, 
\begin{equation}
S_{\rm KW}\,=\,a^{4}\sum_{n,\mu}\bar{\psi}_{n}\gamma_{\mu}{\psi_{n+\mu}-\psi_{n-\mu}\over{2a}}+
a^{5}\sum_{j=1}^{3}\bar{\psi}_{n}i\gamma_{4}{2\psi_{n}-\psi_{n+j}-\psi_{n-j}\over{2a^{2}}},
\label{FCPS}
\end{equation}
\begin{equation}
aD_{\rm KW}(p)=i\gamma_{\mu}\sin p_{\mu}a\,+\,i\gamma_{4}\sum_{j=1}^{3}(1-\cos p_{j} a).
\label{FCPD}
\end{equation}
Here 14 species is decoupled in the naive continuum limit 
while two species at $p=(0,0,0,0)$ and $p=(0,0,0,\pi/a)$
has zero mass and zero imaginary chemical potential
\footnote{These two species are not equivalent since the gamma matrices are 
differently defined between them as $\gamma_\mu'=\Gamma^{-1} \gamma_\mu \Gamma$. 
In the above case it is given by $\Gamma = i \gamma_4 \gamma_5$. 
This means the chiral symmetry possessed by this action is
identified as a flavored one given by $\gamma_5\otimes\tau_3$.}. 
More precisely, among 16 species, two species have zero imaginary 
chemical potential, six have $2/a$, six have $4/a$ and two have $6/a$.
In Fig.~\ref{MD} we compares specie-splitting of KW fermion in chemical potential space
to that of Wilson fermion in mass space.
It is notable that two-flavor is the minimal number allowed by the no-go theorem.
This form has been known as the Karsten-Wilczek (KW) fermion 
\cite{Kar, Wilc}, which is the first known type of ``minimal-doubling fermions"
\cite{Kar, Wilc, Creutz1, Creutz2, Bori, CM}.
It has one exact chiral symmetry, ultra-locality, cubic symmetry, CT and P.
Since the chemical potential term breaks discrete symmetries into
the subgroup, we need to fine-tune three parameters for one dimension-3 
($\bar{\psi}i\gamma_{4}\psi$) and two dimension-4 
($\bar{\psi}\gamma_{4}\partial_{4}\psi$, $F_{j4}F_{j4}$) counterterms in order to take 
a Lorentz-symmetric continuum limit for the zero-($T$,$\mu$) 
lattice QCD simulations \cite{Cap1, Cap2, Cap3}.

Among the three counterterms, in this paper we mainly deal with
the dimension-3 term $\mu_{3}\bar{\psi}i\gamma_{4}\psi$ with a 
relevant parameter $\mu_{3}$ and the dimension-4 term 
$d_{4}\bar{\psi}\gamma_{4}\partial_{4}\psi$ with a marginal parameter $d_{4}$
since we study the strong-coupling lattice QCD and
the Gross-Neveu model, which contain no plaquette action.
In particular the parameter $\mu_{3}$ is of special importance:
It changes the number of flavors and plays an important role in the chiral 
phase structure.
Furthermore the quantum effects produce $O(1/a)$ additive chemical 
potential renormalization in this case instead the additive mass renormalization,
and we need to cancel it by adjusting $\mu_{3}$ even for the application 
to the imaginary-chemical-potential lattice QCD. 
This necessity of parameter tuning is also understood from the well-known fact that 
the naive introduction of chemical potential into lattice fermions leads to 
divergence of energy density and requires a counterterm due to the violation of
the abelian gauge invariance as shown in Ref.~\cite{HK}.

\begin{figure}
\includegraphics[height=7cm]{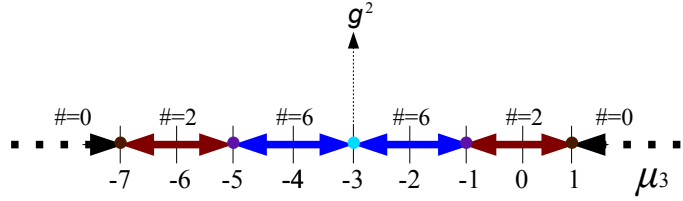} 
\caption{The number of species in Karsten-Wilczek fermion as 
a function of the parameter $\mu_{3}$.
On the boundaries between different sectors, the dispersion relation of
fermions becomes unphysical.}
\label{p-range}
\end{figure}

We here write the KW fermion action of the interacting theory as
\begin{eqnarray}
 S_{\mathrm{KW}} & = &
  \sum_{n} 
  \Bigg[
   \frac{1}{2} \sum_{\mu=1}^4 \bar\psi_n \gamma_\mu
   \left(
    U_{n,n+\mu} \psi_{n+\mu} - U_{n,n-\mu}\psi_{n-\mu}
   \right)
   \nonumber \\
 & & + \frac{r}{2}\sum_{j=1}^3 \bar \psi_n i\gamma_4
  \left( 2\psi_{n} - 
    U_{n,n+j} \psi_{n+j} - U_{n,n-j} \psi_{n-j}
   \right)  + \mu_{3}\bar{\psi}_{n}i\gamma_{4}\psi_{n}+m\bar{\psi}_{n}\psi_{n}
   \nonumber\\
  & & +  {d_{4}\over{2}}\bar\psi_x \gamma_4
   \left(
    U_{n,n+4} \psi_{n+4} - U_{n,n-4}\psi_{n-4}
   \right)
  \Bigg],
\label{Smd}  
\end{eqnarray}
where we introduce a parameter $r$ in analogy with
the Wilson parameter. We introduce the dimension-3 counterterm with
the parameter $\mu_{3}$.
Although we mainly focus on $\mu_{3}$ in this work, 
we also introduce the dimension-4 counterterm with the parameter 
$d_{4}$ in Eq.~(\ref{Smd}) which is also relevant for the strong-coupling study.  
We make all the quantities dimensionless.
Now let us look into how the number of flavors depend on $\mu_{3}$.
For a free theory, the associated massless Dirac operator in momentum 
space is
\begin{equation}
 aD_{\mathrm{KW}}(p) =
  i \sum_{\mu=1}^4 \gamma_\mu \sin ap_\mu
  + i\gamma_4
(\mu_{3}+3r-  r  \sum_{j=1}^{3}\cos ap_j + d_{4}\sin ap_{4}).
 \label{Smdp} 
\end{equation}
We first look into a minimal-doubling range for $\mu_{3}$ and 
the speed of light in the range.
We for simplicity take $r=1$. 
For $-1-d_{4}<\mu_{3}<1+d_{4}$, we have only two zeros, both of which have
the form as 
$\bar{p}=(0, 0, 0, {1\over{a}}\arcsin(-{\mu_{3}\over{1+d_{4}}}))$.
By expanding the momentum as $p=\bar{p}+q$ around the zeros, 
the coefficient of $i\gamma_{4}$ in the Dirac operator is given as
\begin{align}
(1+d_{4})[\sin a\bar{p}_{4}\cdot \cos aq_{4}&+\cos a\bar{p}_{4}\cdot \sin aq_{4}]
+\mu_{3}+3-\sum_{i=1}^{3}\cos (\bar{p}_{i}+q_{i})
\nonumber\\
&=[ (1+d_{4})\cos a\bar{p}_{4}]\, aq_{4} + O(a^{2}q^{2}).
\label{dpeq}
\end{align}
Therefore, for general values of $\mu_{3}$ in this range, the speed of light is modified as
\begin{equation}
\sim q_{4} (1+d_{4})\sqrt{1-{\mu_{3}^{2}\over{(1+d_{4})^{2}}}} +O(aq^{2}).
\label{dp1}
\end{equation}
We can fix this speed of light to a correct value up to the $O(a)$ discretization 
errors by taking $(1+d_{4})^{2}=1+\mu_{3}^{2}$.
We will discuss details of Lorentz symmetry restoration in Sec.\ref{sec:QCD}.
Now, let us classify $\mu_{3}$ parameter regions by the number of physical flavors
in the free theory. Since $d_{4}$ just gives a shift of the parameter regions,
we now consider $d_{4}=0$ for simplicity.
The above minimal-doubling range is given just by $-1<\mu_{3}<1$.   
For $-7<\mu_{3}<-5$, we again have two zeros.
For $\mu_{3}<-7$ and $\mu_{3}>1$, there is no zero of the Dirac operator, 
which means that there are no physical fermions.
For $-5<\mu_{3}<-3$ and $-3<\mu_{3}<-1$, six zeros exist.
$\mu_{3}=1,-7$ have one zero, but
the dispersion relations have a unphysical form as $\sim {\bf p}+p_{4}^{2}$.
$\mu_{3}=-1,-5$ have four zeros, but the dispersions are unphysical.
$\mu_{3}=-3$ has six zeros, but the dispersion is again unphysical.
In the end, the parameter ranges where physical fermions can be described
are $-7<\mu_{3}<-5$ (2 flavors), $-5<\mu_{3}<-3$ (6 flavors),  
$-3<\mu_{3}<-1$ (6 flavors), $-1<\mu_{3}<1$ (2 flavors).
We summarize it in Fig.~\ref{p-range}.
We note that nonzero $d_{4}$ gives larger minimal-doubling range
as shown above. 

What we want to study in this work is how these parameter ranges
evolve in the finite gauge coupling direction.
The most important question for practical use of this formulation in lattice QCD
is how the two-flavor range, or the minimal-doubling range changes in the interacting theory.
Our question is deeply related to a possible chiral phase structure with respect to
the $U(1)$ chiral symmetry. 
It is because spontaneous chiral symmetry breaking is expected to
occur only in parameter ranges with physical fermions.
We thus speculate that boundaries between $\chi$SSB and non-$\chi$SSB phases 
starts from $\mu_{3}=-7$ and $\mu_{3}=1$ in the weak-coupling limit. 
From next section, we will elucidate the $\mu_{3}$-$g^{2}$ chiral phase diagram by using
strong-coupling lattice QCD and the Gross-Neveu model.


\section{Strong-coupling Lattice QCD}
\label{sec:SLQ}

In this section we employ the strong coupling analysis to investigate the chiral
phase structure in lattice QCD with Karsten-Wilczek (KW) fermion. 
The first step is to derive an effective potential of meson fields
corresponding to the fermion action in Eq.~(\ref{Smd}).
The strong-coupling study for KW fermion was first performed by
the present author and collaborators in \cite{Rev}.
We here take the same approach.

Lattice fermion action is generally written as following by using 
hopping operators $P^{\pm}_{\mu}$ and an onsite operator $\hat{M}$  
\begin{equation}
S= \sum_{n,\mu}\bar{\psi}_{n}(P_{\mu}^{+}\psi_{n+\mu}-P_{\mu}^{-}\psi_{n-\mu}) 
+ \sum_{n}\bar{\psi}_{n}\hat{M}\psi_{n}.
\end{equation} 
By using these operators an effective action for mesons in the strong coupling limit can be written 
\cite{KS} as
\begin{eqnarray}
S_{\rm eff}(\mathcal{M}) &=& N_c \sum_n \left[\sum_\mu {\rm Tr}\, f(\Lambda_{n,\mu}) + {\rm tr} \, \hat{M} \mathcal{M}(n) - \tr \, \log \mathcal{M}(n) \right] \, ,\\
\Lambda_{n,\mu}&=&\frac{V_{n,\mu} \bar V_{n,\mu}}{N^2_c} , \quad
\mathcal{M}(n)^{\alpha\beta} =  \frac{\sum_a \bar \psi_n^{a,\alpha}\psi_n^{a,\beta}}{N_c}\, ,
\nonumber
\end{eqnarray}
\begin{eqnarray}
V_{n,\mu}^{ab} &=& \bar\psi_n^b P^-_\mu \psi_{n+\hat\mu}^a\, , \quad
\bar V_{n,\mu}^{ab} = -\bar\psi_{n+\hat\mu}^b P^+_\mu \psi_{n}^a\, , \quad \\
{\rm Tr}\, f(\Lambda_{n,\mu})&=& -{\rm tr}\, f\left( - \mathcal{M}(n) \PpT  \mathcal{M}(n+\hat\mu)\PmT
\right)\,, 
\end{eqnarray}
where $N_c$ is the number of colors, ${\rm Tr}$ ( $\tr$ ) means a trace over color(spinor) index, and $\mathcal{M}(n)$ is a meson field. $a,b$ are indices for colors while $\alpha,\beta$ for spinors.
The explicit form of the function $f$  is determined by performing a one-link integral of the gauge field. 
In the large $N_c$ limit, it is known that $f(x)$ can be analytically evaluated \cite{KaS} as
\begin{equation}
f(x) = \sqrt{1+4x}-1-\ln\frac{1+\sqrt{1+4x}}{2} = x + O(x^2)\, .
\label{eq:largeN}
\end{equation}
For most cases of studying phase structure, we can approximate it as $f(x)\sim x$,
which corresponds to a large-dimension limit \cite{Bl}.
Since the phase transition is expected to be second-order for a massless case,
this approximation at least works well near the phase boundary. 
We however note that it becomes less valid for large $\sigma$.
In the case of the Karsten-Wilczek fermion, we have $\hat M = m {\bf
1}_4 + \iu (\mu_3+3r)\gamma^\T_4$ and
\begin{eqnarray}
P^+_\mu &=& \left\{
\begin{array}{ccc}
 \frac{1}{2} (\gamma_\mu + i  r \gamma_4) & \mu=1,2,3     \\
 \frac{1}{2}\gamma_4 (1+d_{4}) & \mu=4   \\
\end{array}
\right. ,
 \quad
P^-_\mu =
 \left\{
\begin{array}{ccc}
 \frac{1}{2} (\gamma_\mu - i  r \gamma_4) & \mu=1,2,3     \\
 \frac{1}{2}\gamma_4 (1+d_{4}) & \mu=4   \\
\end{array}
\right. .
\end{eqnarray}
We here assume a form of meson condensate with
chiral and 4th vector condensates as
\begin{equation}
\mathcal{M}_0 =\sigma{\bf 1}_4 + i \pi_4\gamma_4.
\end{equation} 
It is because the flavored chemical potential term is expected to
produce 4th vector condensate, which is related to quark density. 
(We will discuss possibility of other condensates in the end of this section.)
The explicit form of the effective action for $\sigma$ and $\pi_{4}$ is
given by
\begin{align}
S_{\rm eff}&=-4N_{c}{\rm Vol.} \mathcal{V}_{\rm eff}(\sigma,\pi_{4}),
\\
\mathcal{V}_{\rm eff}(\sigma,\pi_{4})&=
{1\over{2}}\log (\sigma^{2}+\pi_{4}^{2})-m\sigma +(\mu_{3}+3r)\pi_{4}
\nonumber\\
&\,\,\,\,\,\,\,-{1\over{4}}[3(1+r^2)+(1+d_{4})^{2}]\sigma^{2}
-{1\over{4}}[3(1-r^2)-(1+d_{4})^{2}]\pi_{4}^{2}.
\end{align}
We now find saddle points of $S_{\rm eff}(\mathcal{M})$ from
\begin{align}
{\delta S_{\rm eff}\over{\delta \sigma}}={\delta S_{\rm eff}\over{\delta \pi_{4}}}=0.
\end{align}
Then gap equations are given by
\begin{eqnarray}
\frac{3(1+r^2)+(1+d_{4})^{2}}{2}\sigma + m -\frac{\sigma}{\sigma^2+\pi_4^2} &=& 0 \, ,
\label{g1}
\\
\frac{3(1-r^2)-(1+d_{4})^{2}}{2}\pi_4 -(\mu_3+3r) -\frac{\pi_4}{\sigma^2+\pi_4^2} &=& 0 \,.
\label{g2}
\end{eqnarray}
It is notable that these gap equations have a particle-hole symmetry as 
$(\pi_{4},\mu_{3}+3)\leftrightarrow(-\pi_{4},-\mu_{3}-3)$, 
which is reflected by chiral phase structure as we will see later.
We first consider $m=0$, and solve the equations analytically.
One of the main purposes here is to find a boundary between 
chiral symmetric and broken phases.
For this purpose we take $\sigma=0$ after dividing the first equation by $\sigma$
since $\sigma$ is an order parameter of chiral symmetry breaking.
Then we have
\begin{align}
&{3(1+r^{2})+(1+d_{4})^{2}\over{2}} =\frac{1}{\pi_4^2} \, ,\\
&{3(1-r^{2})-(1+d_{4})^{2}\over{2}}\pi_4 -(\mu_{3}+3r) ={1\over{\pi_{4}}} \,.
\end{align}
These equations give chiral boundaries for $\mu_{3}$ as
\begin{equation}
\mu_{3}=
\pm{6r^{2}+2(1+d_{4})^{2}\over{\sqrt{6r^{2}+2(1+d_{4})^{2}+6}}}-3r.
\label{bound}
\end{equation}
Therefore we have two ranges of $\mu_{3}$ with different chiral properties, {\bf I} and {\bf II} :
\begin{align}
&{\rm {\bf I}}\,:\,\,\,\,\,\,\,\,\,\,\mu_{3}<-{6r^{2}+2(1+d_{4})^2\over{\sqrt{6r^{2}+2(1+d_{4})^{2}+6}}}-3r,\,\,\,\,\,\,\,\,\,\, \mu_{3}>{6r^{2}+2(1+d_{4})^2\over{\sqrt{6r^{2}+2(1+d_{4})^{2}+6}}}-3r, \\
&{\rm {\bf II}}\,:\,\,\,\,\,\,\,\,\,\,-{6r^{2}+2(1+d_{4})^2\over{\sqrt{6r^{2}+2(1+d_{4})^{2}+6}}}-3r
<\mu_{3}<{6r^{2}+2(1+d_{4})^2\over{\sqrt{6r^{2}+2(1+d_{4})^{2}+6}}}-3r. 
\end{align}
\begin{figure}
\includegraphics[height=8cm]{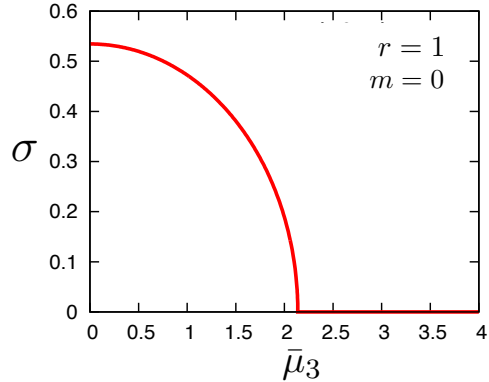} 
\caption{Chiral condensate in strong-coupling lattice QCD with KW fermion
for $m=0$ and $r=1$. We define $\bar{\mu}_{3}\equiv \mu_{3}+3$ and depict the
chiral condensate as a function of $\bar{\mu}_{3}$ for $0<\bar{\mu}_{3}<4$.
This result indicates the second-order chiral transition although the approximation 
$f(x)\sim x$ becomes less reliable for larger $\sigma$.}
\label{chicon}
\end{figure}
The question is which corresponds to the chiral symmetric or broken phases.
We for a while look into the $r=1$ case to show details of chiral phase structure. 
Since the change of $d_{4}$ just shift the chiral phase boundary (\ref{bound}), 
we for a while take $d_{4}=0$ to catch 
the chiral phase structure although we note that we need to take care of it
for a Lorentz symmetric continuum limit.
For $r=1$ and $d_{4}=0$ the boundaries are given by 
$\bar{\mu}_{3}\equiv\mu_{3}+3=\pm\sqrt{32/7}$ ($\mu_{3}\sim-5.14,-0.86$). 
Here we defined shifted $\mu_{3}$ as $\bar{\mu}_{3}=\mu_{3}+3$.
By solving the gap equations in Eq.~(\ref{g1})(\ref{g2}) for $r=1$ and $d_{4}=0$ 
we derive two solutions of chiral and $\pi_{4}$ condensates as
\begin{equation}
\mathcal{M}_{0}^{A}:\,\,\,\,\,\,\,\sigma = 0, \,\,\,\,
\pi_{4}=-\bar{\mu}_{3}+\sqrt{\bar{\mu}_{3}^{2}-2},
\label{chis}
\end{equation}
and
\begin{equation}
\mathcal{M}_{0}^{B}:\,\,\,\,\,\,\,\sigma={\sqrt{32-7\bar{\mu}_{3}^{2}}\over{4\sqrt{7}}},\,\,\,
\pi_{4}=-{\bar{\mu}_{3}\over{4}}.
\label{chib}
\end{equation}
By comparing effective potentials of the two solutions, we find which corresponds to
vacua for the two parameter ranges, {\bf I} and {\bf II}.
We show the following by substituting the two solutions:
$\mathcal{V}_{\mathrm{eff}}(\mathcal{M}_{0}^{A})
-\mathcal{V}_{\mathrm{eff}}(\mathcal{M}_{0}^{B})>0$
for $-\sqrt{32/7}<\bar{\mu}_{3}<\sqrt{32/7}$ ({\bf II})
while $\mathcal{V}_{\mathrm{eff}}(\mathcal{M}_{0}^{A})-\mathcal{V}_{\mathrm{eff}}(\mathcal{M}_{0}^{B})<0$
for $\bar{\mu}_{3}<-\sqrt{32/7},\,\,\bar{\mu}_{3}>\sqrt{32/7}$ ({\bf I}).
To sum up, the chiral symmetric and broken phases are given by
\begin{align}
&\sigma = 0\,\,\,\,\,\,\,\,\,\,\,\,\,\,\,\,\,\,\,\,\,\,\,\,\,\,\,\,\,\,\,\,\,\,\,\,\,\,\,\,\,\,\,\,\,\,\,{\rm for} \,\,\,\,\,\,
\mu_{3}<-\sqrt{32/7}-3,\,\,\mu_{3}>\sqrt{32/7}-3 \,\,\,\,({\rm {\bf I}}),
\\
&\sigma={\sqrt{32-7(\mu_{3}+3)^{2}}\over{4\sqrt{7}}}\,\,\,\,\,{\rm for} \,\,\,\,-\sqrt{32/7}-3<\mu_{3}<\sqrt{32/7}-3 \,\,\,\,({\rm {\bf II}}).
\end{align}
The behavior of chiral condensate is depicted as a function of $\bar{\mu}_{3}=\mu_{3}+3$ 
in Fig.~\ref{chicon}, which shows that the transition 
between chiral symmetric and broken phases is second-order.
We also depict parameter regions for the two phases in the strong-coupling limit 
for $r=1$, $d_{4}=0$ and $m=0$ in Fig.~\ref{stP}.
As seen from Eq.~(\ref{bound}), a nonzero $d_{4}$ gives a larger physical range ({\bf II})
with SSB of chiral symmetry.

\begin{figure}
\includegraphics[height=5cm]{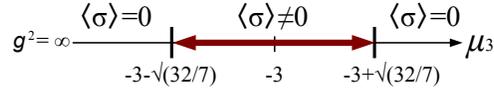} 
\caption{Chiral phase structure in the strong-coupling limit of 
lattice QCD with KW fermion for $r=1$, $d_{4}=0$ and $m=0$.}
\label{stP}
\end{figure}

\begin{figure}
\includegraphics[height=7cm]{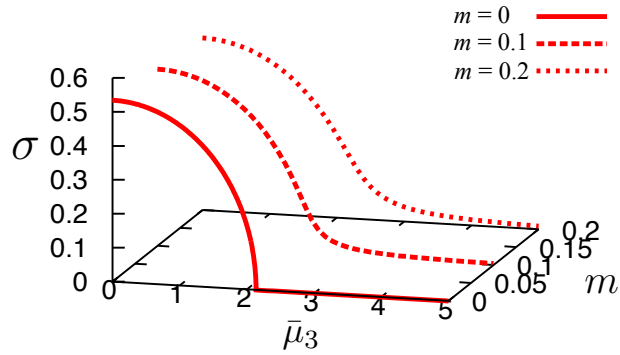} 
\caption{Chiral condensate in strong-coupling lattice QCD with KW fermion
for $r=1$ and $d_{4}=0$ as a function of $\bar{\mu}_{3}$ and $m$.}
\label{try}
\end{figure}

We show that the order of transition becomes a crossover for $m\not=0$.
Fig.\ref{try} shows the chiral condensate as a function of $\bar{\mu}_{3}=\mu_{3}+3$ 
and $m$. It is obvious that the second-order phase transition changes into a crossover
for $m\not=0$.
We note that $\pi_{4}$ condensate is in general non-zero in the chiral-broken phase.
$\pi_{4}$ becomes zero only for $\mu_{3}+3=0$, which corresponds to a theory 
with unphysical dispersions. 

We next look into mass of mesons within the same framework.
We here consider $m=0$ but general values of $r$ and $d_{4}$.
To calculate meson masses we expand the meson field as, 
\begin{equation}
\mathcal{M}(n) = \mathcal{M}^\T_0 + \sum_\A \pi^\A(n) \Gamma^\T_\A \,,\quad
\A \in \left\{ S, P, V_\alpha, A_\alpha, T_{\alpha\beta}\right\}\,,
\label{sectors}
\end{equation}
where $S,P, V_{\alpha},A_{\alpha}$ and $T_{\alpha\beta}$ stand for scalar, pseudo-scalar, vector, axial-vector and tensor respectively.
Here $\mathcal{M}_0$ is the vacuum expectation value of $\mathcal{M}(n)$.
We note that
\begin{eqnarray}
\Gamma_S =\frac{{\bf 1}_4}{2}, \ \Gamma_P =\frac{\gamma_5}{2}, \ \Gamma_{V_\alpha}=\frac{\gamma_\alpha}{2},  \ \Gamma_{A_\alpha}=\frac{i \gamma_5\gamma_\alpha}{2}, \
\Gamma_{T_{\alpha\beta}}=\frac{\gamma_\alpha\gamma_\beta}{2i }\ (\alpha < \beta).
\end{eqnarray}
Then the effective action at the second order of $\pi^\A$ is given by
\begin{eqnarray}
S_{\rm eff}^{(2)} &=& N_c \sum_n\biggl[
\frac{1}{2}{\rm tr}\,( \mathcal{M}_0^{-1} \Gamma_\A \mathcal{M}_0^{-1}\Gamma_\B) \,  \pi^\A(n)\pi^\B(n) 
+ \sum_\mu {\rm tr}\, ( \Gamma_\A P^{-}_\mu \Gamma_\B P^{+}_\mu ) \pi^\A(n) \pi^\B(n+\hat\mu)
\biggr] \nonumber \\
&=& N_c\int\frac{d^4 p}{(2\pi)^4} \pi^\A(-p) D_{\A\B}(p) \pi^\B(p) \, ,
\end{eqnarray}
where
\begin{eqnarray}
D_{\A\B} (p) &=& \frac{1}{2} \bigl( \widetilde{D}_{\A\B}(p) + \widetilde{D}_{\B\A} (-p)\bigr),\\[.5ex]
\widetilde{D}_{\A\B}(p) &=& \frac{1}{2}{\rm tr}\,(\mathcal{M}_0^{-1} \Gamma_\A \mathcal{M}_0^{-1}\Gamma_\B)
+ \sum_\mu {\rm tr}\, ( \Gamma_\A P^-_\mu \Gamma_\B P^+_\mu ) e^{i p_\mu} .
\end{eqnarray}
In our case $\mathcal{M}_{0}=\sigma{\bf 1}+i\pi_{4}\gamma_{4}$ gives
\begin{equation}
\mathcal{M}_{0}^{-1}={1\over{\sigma^{2}+\pi_{4}^{2}}}(\sigma{\bf 1}-i\pi_{4}\gamma_{4}).
\end{equation} 
We now write the whole inverse meson propagator matrix 
in the $S$-$V$-$T$-$A$-$P$ sector as
\begin{align}
&D_{SVTAP}=
\nonumber\\ 
\nonumber\\
&\left[
\begin{array}{cccccccccccccccc}
D_{S} & -C &  {-irs_{1}\over{2}} & {-irs_{2}\over{2}} & {-irs_{3}\over{2}} &  & & & & & & & & & & \\
-C & D_{V_{4}} & & & & & & & & & & & & & & \\
{irs_{1}\over{2}}& & D_{T_{14}} & & &  & & & & & &  &  & & & \\
{irs_{2}\over{2}}& & & D_{T_{24}} & & & & & & & &  & & & & \\
{irs_{3}\over{2}}& & & & D_{T_{34}} & & & & & & & & & & & \\
& & & &  & D_{V_{3}} & & & & & &  {-irs_{2}\over{2}}  &  {irs_{1}\over{2}}  & & & \\
& & & & & & D_{V_{2}} & & & & &  {irs_{3}\over{2}}   & &  {-irs_{1}\over{2}}  & & \\
& & & & & & & D_{V_{1}} & & & & & {-irs_{3}\over{2}}    &   {irs_{2}\over{2}}  & & \\
& & & & & & & & D_{T_{12}} &  &  & &  & C &  {-irs_{3}\over{2}} &     \\
& & & & & & & &  & D_{T_{13}} &  &  & -C & & {irs_{2}\over{2}} & \\
& & & & & & & &  &  & D_{T_{23}} & C &  & & {-irs_{1}\over{2}} & \\
& & & & & {irs_{2}\over{2}} &  {-irs_{3}\over{2}} & &  &  & C & D_{A_{1}} &  &  &  \\
& & & & &  {-irs_{1}\over{2}} & & {irs_{3}\over{2}} & & -C &  &  & D_{A_{2}} &  &  \\
& & & & & &{irs_{1}\over{2}} & {-irs_{2}\over{2}} & C &  &  &  &  & D_{A_{3}} &  \\
& & & & & & & & {irs_{3}\over{2}} & {-irs_{2}\over{2}} & {irs_{1}\over{2}} & & & & D_{P}  & \\
& & & & & & & &  & & &  &  & & & D_{A_4} \\
\end{array}
\right] \, ,
\nonumber\\
\end{align}
where components are given by
\begin{align}
D_{S}&= {\sigma^{2}-\pi_{4}^{2}\over{2(\sigma^{2}+\pi_{4}^{2})^{2}}}
+ {1\over{4}}[ (1+r^{2})(c_{1}+c_{2}+c_{3}) + (1+d_{4})^{2}c_{4}], 
\label{Ds}
\\
D_{V_{4}}&={\sigma^{2}-\pi_{4}^{2}\over{2(\sigma^{2}+\pi_{4}^{2})^{2}}}
-{1\over{4}}[(1-r^{2})(c_{1}+c_{2}+c_{3})-(1+d_{4})^{2}c_{4}],\\
D_{V_{3}}&= {1\over{2(\sigma^{2}+\pi_{4}^{2})}}
- {1\over{4}}[ (1+r^{2})(c_{1}+c_{2})+(r^{2}-1)c_{3} + (1+d_{4})^{2}c_{4} ], \\
D_{V_{2}}&= {1\over{2(\sigma^{2}+\pi_{4}^{2})}}
- {1\over{4}}[ (1+r^{2})(c_{1}+c_{3})+(r^{2}-1)c_{2} + (1+d_{4})^{2}c_{4} ], \\
D_{V_{1}}&= {1\over{2(\sigma^{2}+\pi_{4}^{2})}}
- {1\over{4}}[(1+r^{2})(c_{2}+c_{3}) +  (r^{2}-1)c_{1}+(1+d_{4})^{2}c_{4} ], \\
D_{T_{14}}&= {1\over{2(\sigma^{2}+\pi_{4}^{2})}}
- {1\over{4}}[ (r^{2}-1)(c_{2}+c_{3})+(1+r^{2})c_{1} + (1+d_{4})^{2}c_{4} ], \\
D_{T_{24}}&= {1\over{2(\sigma^{2}+\pi_{4}^{2})}}
- {1\over{4}}[ (r^{2}-1)(c_{1}+c_{3})+(1+r^{2})c_{2} + (1+d_{4})^{2}c_{4} ], \\
D_{T_{34}}&= {1\over{2(\sigma^{2}+\pi_{4}^{2})}}
- {1\over{4}}[ (r^{2}-1)(c_{1}+c_{2})+(1+r^{2})c_{3} + (1+d_{4})^{2}c_{4} ], \\
D_{T_{12}}&= {\sigma^{2}-\pi_{4}^{2}\over{2(\sigma^{2}+\pi_{4}^{2})^{2}}}
+ {1\over{4}}[ (r^{2}-1)(c_{1}+c_{2})+(1+r^{2})c_{3} + (1+d_{4})^{2}c_{4} ], \\
D_{T_{13}}&= {\sigma^{2}-\pi_{4}^{2}\over{2(\sigma^{2}+\pi_{4}^{2})^{2}}}
+ {1\over{4}}[ (r^{2}-1)(c_{1}+c_{3})+(1+r^{2})c_{2} + (1+d_{4})^{2}c_{4} ], \\
D_{T_{23}}&= {\sigma^{2}-\pi_{4}^{2}\over{2(\sigma^{2}+\pi_{4}^{2})^{2}}}
+ {1\over{4}}[ (r^{2}-1)(c_{2}+c_{3})+(1+r^{2})c_{1} + (1+d_{4})^{2}c_{4} ], \\
D_{A_{1}}&= {\sigma^{2}-\pi_{4}^{2}\over{2(\sigma^{2}+\pi_{4}^{2})^{2}}}
+ {1\over{4}}[(1+r^{2})(c_{2}+c_{3}) +  (r^{2}-1)c_{1}+(1+d_{4})^{2}c_{4} ], \\
D_{A_{2}}&= {\sigma^{2}-\pi_{4}^{2}\over{2(\sigma^{2}+\pi_{4}^{2})^{2}}}
+ {1\over{4}}[ (1+r^{2})(c_{1}+c_{3})+(r^{2}-1)c_{2} + (1+d_{4})^{2}c_{4} ], \\
D_{A_{3}}&= {\sigma^{2}-\pi_{4}^{2}\over{2(\sigma^{2}+\pi_{4}^{2})^{2}}}
+ {1\over{4}}[ (1+r^{2})(c_{1}+c_{2})+(r^{2}-1)c_{3} + (1+d_{4})^{2}c_{4} ], \\
D_{A_{4}}&={1\over{2(\sigma^{2}+\pi_{4}^{2})}}
+{1\over{4}}[(1-r^{2})(c_{1}+c_{2}+c_{3})-(1+d_{4})^{2}c_{4}],
\label{Da4}
\\
D_{P}&= {1\over{2(\sigma^{2}+\pi_{4}^{2})}}
- {1\over{4}}[ (1+r^{2})(c_{1}+c_{2}+c_{3}) + (1+d_{4})^{2}c_{4}], \\
C&= {i\sigma\pi_{4}\over{(\sigma^{2}+\pi_{4}^{2})^{2}}},
\end{align}
with $s_k =\sin p_k$ and $c_k=\cos p_k$.
By diagonalizing this matrix, we can derive 
an explicit form of physical meson propagators.
We note that $C$ gets zero in the chiral symmetric phase
while it has nonzero values in the chiral broken phase.

We first check that pion mass becomes zero in the chiral-broken phase.
For this purpose we substitute $p=(0,0,0, i m_P)$ into the propagator.
Then $D_{P}$ is decoupled and the calculation gets simplified.
We note that, in the chiral broken phase, the gap equation Eq.~(\ref{g1}) 
gives $\sigma^{2}+\pi_{4}^{2} =2/[3(1+r^{2})+(1+d_{4})^{2}]$.
Pion mass in the chiral broken phase with $r=1$ and $m=0$ is 
derived from a pole of $D_{P}$ with $p=(0,0,0, i  m_P)$ as
\begin{align}
D_{P}(0,0,0,im_{P})= {3(1+r^{2})+(1+d_{4})^{2}\over{4}}-{1\over{4}}((1&+d_{4})^{2}\cos(i m_{P})+3(1+r^{2})) =0
\nonumber\\
&\,\,\,\,\,\,\,\,\,\,\,\to\,\,\,\,\,\,\,\,\,\,\, \cosh(m_P) = 1 .
\end{align}
This mode corresponds to a massless NG boson associated with SSB of 
chiral $U(1)$ symmetry, which is consistent with realistic QCD.

On the second-order phase boundary, a divergent correlation length 
in the critical behavior should produce another massless mode.
We here show that the scalar meson becomes massless on the critical line.
An inverse meson propagator matrix in general has off-diagonal components, 
but substitution of $p=(0,0,0, i  m_S)$ again decouples the scalar sector 
in the chiral-symmetric phase ($C=0$).
An explicit form of the scalar inverse propagator is given by Eq.~(\ref{Ds}).
Then we derive the scalar mass from a pole of this propagator with 
$p=(0,0,0, i  m_S)$ as
\begin{equation}
D_{S}(0,0,0,im_{S})=0\,\,\,\,\,\to\,\,\,\,
\cosh m_{S} =1-\Big[ 1+{3(1+r^{2})\over{(1+d_{4})^{2}}}+{1\over{(1+d_{4})^{2}}}{2(\sigma^{2}-\pi_{4}^{2})\over{(\sigma^{2}+\pi_{4}^{2})^{2}}} \Big] .
\end{equation}
As seen from this, scalar meson has nonzero mass in general.
On the phase boundary, however, we have $\sigma=0$, 
$\pi_{4} =-\bar{\mu}_{3}+\sqrt{\bar{\mu}_{3}^{2}+2[3(1-r^{2})-(1+d_{4})^{2}]}$ and 
$\bar{\mu}_{3}=\pm{6r^{2}+2(1+d_{4})^{2}\over{\sqrt{6r^{2}+2(1+d_{4})^{2}+6}}}$
with $\bar{\mu}_{3}=\mu_{3}+3r$.
It leads to zero $\sigma$ mass as
\begin{equation}
\cosh(m_S) = 1 .
\end{equation}
We can also show this from the chiral-broken phase.
We have two massless modes only on the 2nd-order phase boundary,
$\sigma$-meson and $\pi$-meson, which is inconsistent with the meson mass 
spectrum in QCD. We thus need to avoid this point in the simulation of QCD.
In the first place this boundary corresponds to the boundary between
the two-flavor and no-flavor ranges, which has the unphysical dispersion relation
as shown in Sec.~\ref{sec:FCP}.
From the theoretical viewpoint, it is however an attractive topic.
Further study including numerical simulations can
elucidate detailed properties of this point.

Now let us discuss possibility of restoration of Lorentz symmetry.
In the strong-coupling limit, what we can do is 
tuning of $\mu_{3}$ and $d_{4}$
since we can ignore tuning for the plaquette action in this limit.
However, we cannot restore the Lorentz symmetry correctly
in this limit because of the large lattice artifacts. 
The $A_{4}$ meson propagator explicitly manifests this point as following:
The $A_{4}$-sector propagator,
which is diagonal even with finite spacial momentum, 
is given by Eq.~(\ref{Da4}).
In the chiral-broken phase, we have $1/(\sigma^{2}+\pi_{4}^{2})=[3(1+r^{2})+(1+d_{4})^{2}]/2$.
By substituting $p=(0,0,0,im_{A4})$, the pole of the propagator determines 
the axial vector meson mass as
\begin{equation}
\cosh m_{A_4}=1+{6\over{(1+d_{4})^{2}}}.
\end{equation}
For small but finite spatial momentum as $p=(p_{1},p_{2},p_{3},iE)$,
the pole of the propagator up to $O(p^{2})$ gives
\begin{equation}
E^{2}=m_{A_4}^{2}+{1-r^{2}\over{(1+d_{4})^{2}}}{\bf p}^{2}.
\label{a4d}
\end{equation} 
It apparently indicates that we need to tune $d_{4}$ as $(1+d_{4})^{2}=1-r^{2}$.
However, for $r=1$, the dependence on ${\bf p}$ itself disappears
and the Lorentz symmetry cannot be restored.
It is a typical strong-coupling artifact. 
Moreover, the tuned value of $d_{4}$ does not depend on $\mu_{3}$, 
but it is also a lattice artifact.
In the first place, the equation (\ref{g1}) indicates that $\pi_{4}$ always has a nonzero value 
in the physical parameter range ($\sigma\not=0$) even if we tune $\mu_{3}$ and $d_{4}$ 
independently except for $\mu_{3}+3r=0$.
All these results show that it is difficult to restore the symmetry in the physical 
parameter range within the framework of the strong-coupling QCD.
We expect that it is just a strong-coupling artifact, and
we can make $\pi_{4}$ zero and restore Lorentz symmetry in the
weak coupling by tuning the three parameters appropriately.

In this section we have assumed the form of condensation $\sigma+i\gamma_{4}\pi_{4}$.
As shown in the Appendix.A, we can also consider possibility of other condensations
as $\sigma+i\gamma_{4}\pi_{4}+i\gamma_{5}\pi_{5}$ or 
$\sigma+i\gamma_{4}\pi_{4}+i\gamma_{4}\gamma_{5}\pi_{45}$. 
The results show that the solution
with nonzero $\pi_{5}$ or $\pi_{45}$ condensates 
cannot be a vacuum, and our solutions in this section
are likely to be true vacua.
We thus consider that the parity breaking phase will not
appear in the KW fermion unless we introduce the flavored-mass terms shown in \cite{CKM1}.
At the weak coupling there may be a more subtle competition between the discretization
error and the counterterm.
We thus need further study to conclude whether the parity breaking
exist or not when we take into account all the three counterterms 
in the weak coupling.

In the end of this section, we discuss the other type of minimal-doubling fermions,
called the Creutz-Borici type \cite{Bori, Creutz1}. 
We can analyze it in a parallel way.
We note that this type specifies the diagonal direction characterized by 
$2\Gamma=\gamma_{1}+\gamma_{2}+\gamma_{3}+\gamma_{4}$, 
instead of the time direction.
Appendix.~B is devoted to detailed analysis for this case.
The result is qualitatively the same.
We find two chiral boundaries and 
chiral condensate is nonzero between the boundaries.


\section{Gross-Neveu model}
\label{sec:GN}

We investigate the whole phase diagram for Karsten-Wilczek (KW)
fermion by using the two-dimensional lattice Gross-Neveu model
\cite{GN, EN, AH, Leder1, Korzec, Leder2, TNU, TN, CKM2, Kama}, 
which has common features with 4d lattice QCD. 
In two dimensions, massless KW fermion action is given by
\begin{equation}
 aD_{\mathrm{KW}}(p) =
  i \sum_{\nu=1}^2 \gamma_\nu \sin ap_\nu
  + i\gamma_2
[r(1-  \cos ap_1)+\mu+d \sin ap_{2}].
 \label{2dD} 
\end{equation}
In this section we concentrate on the case with $r=1$.
In this section we denote the relevant parameter
$\mu_{3}$ as just $\mu\equiv\mu_{3}$
and denote the maginal parameter $d_{4}$ as $d\equiv d_{4}$.
To look into the number of flavors in a free theory,
we for a while consider $d=0$. 
For $-3<\mu<-1$ and $-1<\mu<1$,
there are only two zeros, and it becomes
minimal-doubling.
For $\mu<-3$ and $\mu>1$,
there is no zero, and it becomes a fermion-less
theory.
For $\mu=-3, 1$, there is one zero, but the dispersion relation
becomes unphysical $\sim p_{1}+p^{2}_{2}$.
For $\mu=-1$, there two zeros but whose dispersion relation
is again unphysical.
The main difference from four-dimensional cases is that 
there is no 6-flavor range.

The lattice Gross-Neveu model with KW fermion is given by
\begin{align}
S\,=\,
{1\over{2}}\sum_{n,\nu}\bar{\psi}_{n}&\gamma_{\nu}(\psi_{n+\nu}-\psi_{n-\nu})
+{1\over{2}}\sum_{n}\bar{\psi}_{n}i\gamma_{2}(2\psi_{n}-\psi_{n+\hat{1}}-\psi_{n-\hat{1}})
\nonumber\\
&-{1\over{2N}}\sum_{n}[g_{\sigma}^{2}(\bar{\psi}_{n}\psi_{n})^{2}
+g_{2}^{2}(\bar{\psi}_{n}i\gamma_{2}\psi_{n})^{2}]
\nonumber\\
&+\mu\sum_{n}\bar{\psi}_{n}i\gamma_{2}\psi_{n}
+d\sum_{n}\bar{\psi}_{n}i\gamma_{2}(\psi_{n+\hat{2}}-\psi_{n-\hat{2}})
+m\sum_{n}\bar{\psi}_{n}\psi_{n},
\label{SGNS}
\end{align}
where $\nu$ stands for $\nu=1,2$, $n=(n_1, n_2)$ are the two dimensional 
coordinates and $\psi_{n}$ stands for a $N$-component Dirac fermion field
$(\psi_{n})_{j}$($j=1,2,...,N$). 
The bilinear $\bar{\psi}\psi$ means $\sum_{j=1}^{N}\bar{\psi}_{j}\psi_{j}$, and 
($g_{\sigma}^{2}$, $g_{2}^{2}$) corresponds to the 't Hooft couplings
for the two types of four-fermi interactions.  
Here we define the two dimensional gamma matrices as
$\gamma_{1}=\sigma_{1}$, $\gamma_{2}=\sigma_{2}$ and
$\gamma_{3}=\sigma_{3}$.  We make all the quantities dimensionless in
this equation. 
We consider scalar and time-direction vector four-fermi 
interactions, which are natural choices since KW fermion specifies
the time direction.
We note that, if we drop a mass term ($m=0$), 
the action has $\pi/2$ discrete chiral symmetry,
which can be spontaneously broken due to chiral
condensate.
By introducing auxiliary bosonic fields $\sigma(n)$,
$\pi_{2}(n)$ we remove the four-point interactions as
\begin{align}
S\,=\, {1\over{2}}\sum_{n,\nu}\bar{\psi}_{n}&\gamma_{\nu}(\psi_{n+\nu}-\psi_{n-\nu})
-\sum_{n}\bar{\psi}_{n}i\gamma_{2}(\psi_{n+1}+\psi_{n-1})
+d\sum_{n}\bar{\psi}_{n}i\gamma_{2}(\psi_{n+\hat{2}}-\psi_{n-\hat{2}})
\nonumber\\
&+{N\over{2}}\sum_{n}[{1\over{g_{\sigma}^{2}}}(\sigma(n)-m)^{2}+{1\over{g_{2}^{2}}}(\pi_{2}(n)-\mu-1)^{2}]
+\sum_{n}\bar{\psi}_{n}[\sigma(n)+i\gamma_{2}\pi_{2}(n)]\psi_{n}.
\label{SGNSsp}
\end{align}
By solving the equations of motion, we show the following relation between these
auxiliary fields and the bilinears of the fermion fields
\begin{align}
\sigma(n)&=m-{g_{\sigma}^{2}\over{N}}\bar{\psi}\psi,
\label{sigma}
\\
\pi_{2}(n)&=1+\mu-{g_{2}^{2}\over{N}}\bar{\psi}i\gamma_{2}\psi.
\label{pi}
\end{align}
These relations indicate that $\sigma$ and $\pi_{2}$ stand for the scalar
and vector mesons.  After integrating the fermion fields, the
partition function and the effective action with these auxiliary
fields are given by
\begin{align}
Z\,&=\, \int\prod_{n}d\sigma(n)d\pi_{2}(n)e^{-N\,S_{\rm eff}(\sigma,\pi_{2})},
\label{Par}
\\
S_{\rm eff}(\sigma,\pi_{2}) \,&=\, {1\over{2}}\sum_{n}[{1\over{g_{\sigma}^{2}}}(\sigma(n)-m)^{2}
+{1\over{g_{2}^{2}}}(\pi_{2}(n)-\mu-1)^{2}]-{\rm Tr}\,\log D_{n,m},
\label{NSeff}
\end{align}
with
\begin{align}
D_{n,m}=[\sigma(n)+i\gamma_{2}\pi_{2}(n)]\delta_{n.m}&+
{\gamma_{\mu}\over{2}}(\delta_{n+\mu,m}-\delta_{n-\mu,m})
\nonumber\\
&-{i\gamma_{2}\over{2}}(\delta_{n+\hat{1},m}+\delta_{n-\hat{1},m})
+d{\gamma_{2}\over{2}}(\delta_{n+\hat{2},m}-\delta_{n-\hat{2},m}).
\label{ND}
\end{align}
Here ${\rm Tr}$ stands for the trace both for the position and spinor spaces.
As is well-known, the partition function in the Gross-Neveu model is given 
by the saddle point of this effective action in the large $N$ limit.
We denote as $\tilde{\sigma}(n)$, $\tilde{\pi}_{2}(n)$ solutions satisfying 
the saddle-point conditions
\begin{equation}
{\delta S_{\rm eff}[\sigma(n),\pi_{2}(n)]\over{\delta \sigma(n)}}\,=\, 
{\delta S_{\rm eff}[\sigma(n),\pi_{2}(n)]\over{\delta \pi_{2}(n)}}\,=\, 0.
\label{NSadEq1}
\end{equation}
Then the partition function is given by
\begin{equation}
Z\,=\, e^{-NS_{\rm eff}[\tilde{\sigma},\tilde{\pi_{2}}]}.
\label{NeffPar}
\end{equation}
By assuming the translational invariance we define the position-independent solutions 
as $\sigma\equiv\tilde{\sigma}(0)$ and $\pi\equiv\tilde{\pi_{2}}(0)$ 
Then we can factorize a volume factor $V=\sum_{n}1$ in the effective action as 
\begin{align}
S_{\rm eff}\,&=\, V\tilde{S}_{\rm eff}(\sigma,\pi_{2}),
\\
\tilde{S}_{\rm eff}(\sigma,\pi_{2})
\,&=\, {1\over{2g_{\sigma}^{2}}}(\sigma-m)^{2}
+{1\over{2g_{2}^{2}}}(\pi_{2}-\mu-1)^{2}-{1\over{V}}{\rm Tr}\,\log D.
\label{NtilSeff}
\end{align}
We can write ${\rm Tr}\log D$ in a simple form by the Fourier transformation to momentum space
\begin{align}
{\rm Tr}\,\log D \,&=\,V\int{d^{2}k\over{(2\pi)^{2}}}\log[{\rm det}(\sigma+i\gamma_{2}\pi_{2}+
i\gamma_{2}((1+d)\sin k_{2}-\cos k_{1})+i\gamma_{1}\sin k_{1})]
\nonumber\\
&=\, V\int{d^{2}k\over{(2\pi)^{2}}}\log[\sigma^{2}+(\pi_{2}+(1+d)\sin k_{2}-\cos k_{1})^{2}+(\sin k_{1})^{2}],
\label{NI}
\end{align}
with ${\rm det}$ being the determinant in the spinor space.
Now saddle-point equations are written as
\begin{align}
{\delta \tilde{S}_{\rm eff}\over{\delta \sigma}}\,&=\,
{(\sigma-m)\over{g_{\sigma}^{2}}}-2\int{d^{2}k\over{(2\pi)^{2}}}
{\sigma\over{\sigma^{2}+(\pi_{2}+(1+d)\sin k_{2}-\cos k_{1})^{2}+(\sin k_{1})^{2}}}=0,
\label{Ncond1}
\\ 
{\delta \tilde{S}_{\rm eff}\over{\delta \pi_{2}}}\,&=\,
{\pi_{2}-\mu-1\over{g_{2}^{2}}}-2\int{d^{2}k\over{(2\pi)^{2}}}
{\pi_{2}+(1+d)\sin k_{2}-\cos k_{1}\over{\sigma^{2}+(\pi_{2}+(1+d)\sin k_{2}-\cos k_{1})^{2}+(\sin k_{1})^{2}}}=0.
\label{Ncond2}
\end{align}
Here the values of $\sigma$ and $\pi_{2}$ in the vacuum are determined as 
$\sigma(\mu,d,g_{\sigma}^{2},g_{2}^{2},m)$, 
$\pi_{2}(\mu,d,g_{\sigma}^{2},g_{2}^{2},m)$ from the saddle-point equations 
once $\mu$, $d$, $g_{\sigma}^{2}$, $g_{2}^{2}$ and $m$ are fixed. 

\begin{figure}
\includegraphics[height=7cm]{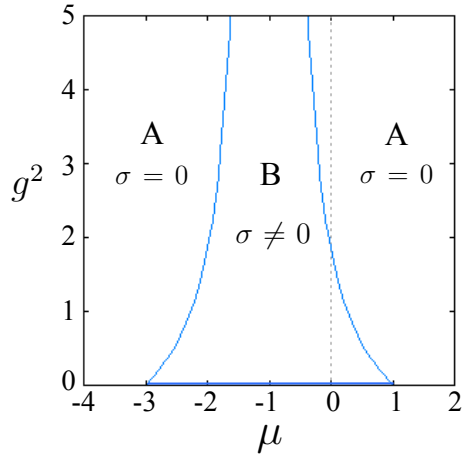} 
\caption{Chiral phase diagram for KW Gross-Neveu model in the $\mu$-$g^{2}$ space.
A stands for the chiral symmetric phase. B stands for the chiral-broken phase. The chiral boundaries
are connected to the edges of minimal-doubling ranges.}
\label{fig1}
\end{figure}

Let us look into the phase diagram with respect to chiral symmetry.
We here consider a massless case as $m=0$ to have the exact discrete 
chiral symmetry in the action.
To capture rough structure of the phase diagram,
we first take the simplest case with $d=0$ and $g_{\sigma}^{2}=g_{2}^{2}\equiv g^{2}$.
Nonzero $d$ just gives slight change of the phase diagram.
Since a single coupling constant works when we study the Aoki phase in 
the Wilson fermion \cite{AokiP}, we expect that the above condition for the
couplings works at least for deriving rough phase structure.  
The order parameter is $\sigma$, which can be zero or non-zero 
depending on values of $\mu$ and $g^{2}$.
The phase boundary is determined by imposing $\sigma=0$ on 
Eq.~(\ref{Ncond1})(\ref{Ncond2}) after the overall $\sigma$ being removed in Eq.~(\ref{Ncond1}).
Then the conditions for the phase boundary are given by gap equations as
\begin{align}
{\pi_{2}-\mu_{c}-1\over{g^{2}}}\,&=\,2\int{d^{2}k\over{(2\pi)^{2}}}
{\pi_{2}+\sin k_{2}-\cos k_{1}\over{(\pi_{2}+\sin k_{2}-\cos k_{1})^{2}+(\sin k_{1})^{2}}},
\label{Ncricon1}
\\
{1\over{g^{2}}}\,&=\,2\int{d^{2}k\over{(2\pi)^{2}}}
{1\over{(\pi_{2}+\sin k_{2}-\cos k_{1})^{2}+(\sin k_{1})^{2}}},
\label{Ncricon2}
\end{align}
with $\mu_{c}$ being a critical value of $\mu$.
Here we derive the chiral phase boundary $\mu_{c}(g^{2})$ as a function of the coupling $g^{2}$
by getting rid of vector condensate $\pi_{2}$ from these equations.
The phase diagram is depicted in Fig.~{\ref{fig1}}.
A stands for the chiral symmetric phase $\sigma=0$ 
and B for chiral broken phase $\sigma\not=0$.
For $-2<\pi_{2}<2$ the integral in Eq.(\ref{Ncricon2}) diverges, which leads to
the critical lines with zero gauge coupling for $-3<\mu<1$ as shown in Fig.~{\ref{fig1}}.
This is reasonable since the weak-coupling limit should have zero chiral condensate.
As we expected, the chiral critical line is connected to boundaries 
between two-flavor and no-flavor phases in the weak-coupling limit ($\mu=-3,1$).
It is consistent with our intuition that the fermion-less theory cannot cause spontaneous
breaking of chiral symmetry.  
We can check the mass of the scalar meson mass becomes 
zero on the critical line $\mu_{c}(g^{2})$.
The mass of $\sigma$ is calculated analytically on the boundary, and is shown to be zero as
\begin{align}
m_{\sigma}^{2}\, &\propto\, 
\langle {\delta^{2}S_{\rm eff}\over{\delta \sigma(n) \delta \sigma(m)}} \rangle_{\mu_{c}} 
=V{\delta^{2}\tilde{S}_{\rm eff}\over{\delta^{2}\sigma^{2}}}|_{\mu_{c}}
\nonumber\\
&=V\Big[ {1\over{g^{2}}}
-2\int{d^{2}k\over{(2\pi)^{2}}}{1\over{\sigma^{2}+(\pi_{2}+\sin k_{2}-\cos k_{1})^{2}+(\sin k_{1})^{2}}}
\nonumber\\
&\,\,\,\,\,\,\,\,\,\,\,\,\,\,\,\,\,\,\,\,\,\,\,-4\sigma^{2}\int{d^{2}k\over{(2\pi)^{2}}}{1\over{\sigma^{2}+(\pi_{2}+\sin k_{2}-\cos k_{1})^{2}+(\sin k_{1})^{2}}} \Big]_{\mu_{c}}
\nonumber\\
&=0.
\end{align}
A massless scalar meson indicates that the phase boundary we derived is a second-order
critical line. This result is consistent with the strong-coupling lattice QCD 
in the previous section.

\begin{figure}
\includegraphics[height=7cm]{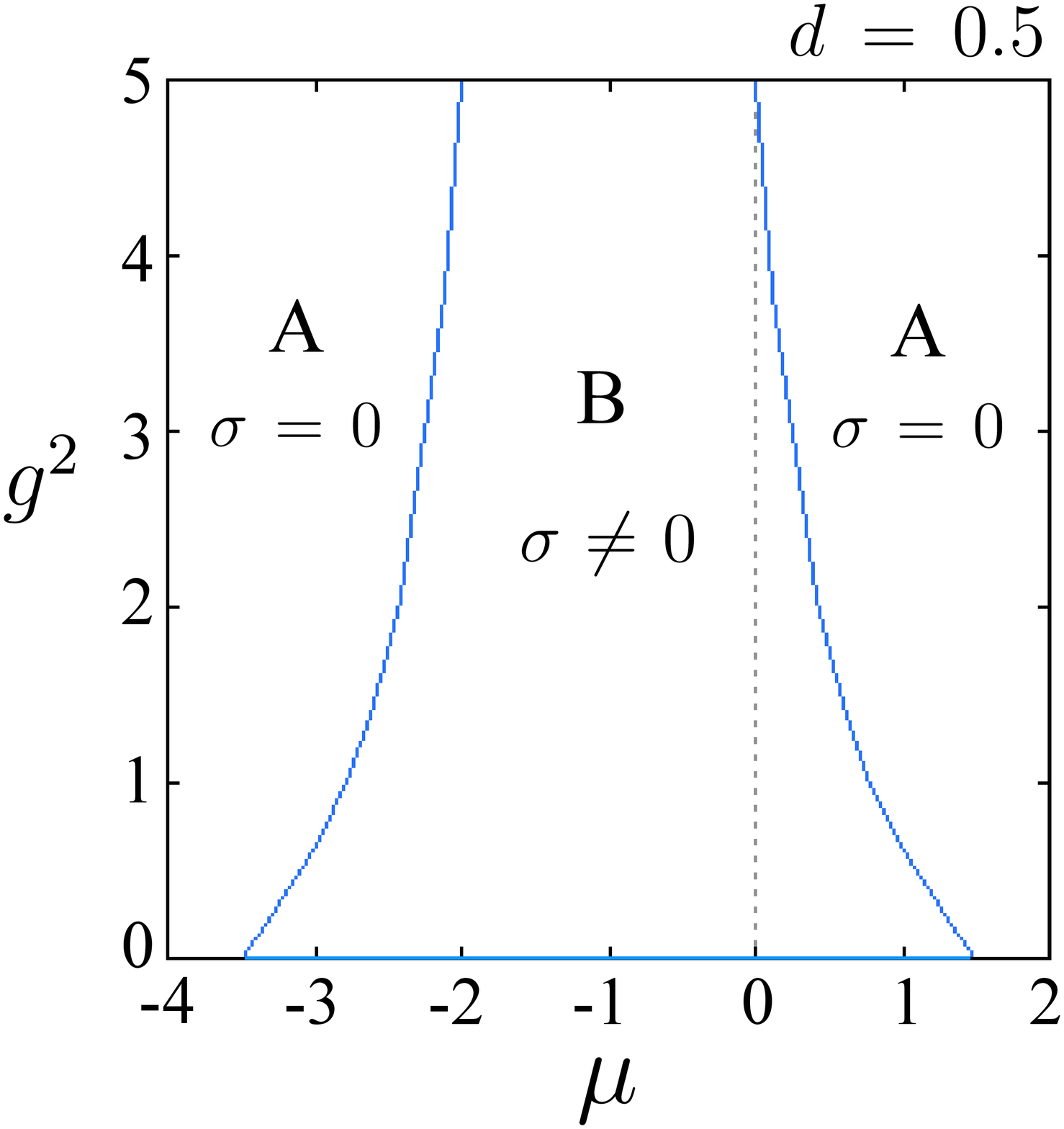} 
\caption{Chiral phase diagram for KW Gross-Neveu model in the $\mu$-$g^{2}$ space
with $d=0.5$ ($g^{2}\equiv g_{\sigma}^{2}=g_{2}^{2}$). The physical (minimal-doubling) phase with chiral condensate is enlarged by nonzero values of $d$.In the weak-coupling limit, the minimal-doubling phases are given by $-3.5<\mu<-1$ and $-1<\mu<1.5$.}
\label{0.5}
\end{figure}

Next, we discuss more general cases with $d\not=0$ and $g_{\sigma}^{2}\not= g_{2}^{2}$.
Nonzero values of $d$ change $\sin k_{2}\to(1+d)\sin k_{2}$ 
in (\ref{Ncricon1})(\ref{Ncricon2}) and give just qualitative changes of the phase diagram:
Since the minimal-doubling ranges are given by $-1<\mu<1+d$ and
$-3-d<\mu<-1$ for nonzero $d$, it gives a larger physical range in the phase diagram.
As an example, we depict the $\mu$-$g^{2}$ phase diagram for $d=0.5$ in Fig.~\ref{0.5}.
In the case with two independent coupling constants $g_{\sigma}^{2}\not= g_{2}^{2}$,
the equations for the phase boundary are given by
\begin{align}
\mu_{c}+1\,&=\,\pi_{2}\left( 1-{g_{2}^{2}\over{g_{\sigma}^{2}}} \right)-2g_{2}^{2}\int{d^{2}k\over{(2\pi)^{2}}}
{\pi_{2}+\sin k_{2}-\cos k_{1}\over{(\pi_{2}+\sin k_{2}-\cos k_{1})^{2}+(\sin k_{1})^{2}}},
\label{1con1}
\\
{1\over{g_{\sigma}^{2}}}\,&=\,2\int{d^{2}k\over{(2\pi)^{2}}}
{1\over{(\pi_{2}+\sin k_{2}-\cos k_{1})^{2}+(\sin k_{1})^{2}}}.
\label{1con2}
\end{align}
In this case the phase diagram is deformed to some extent:
If we fix one of the coupling constants and depict the phase diagram, 
the physical phase gets larger toward the strong-coupling limit while it gets narrower 
at the weak-coupling. As an example, we depict the $\mu$-$g_{\sigma}^{2}$ 
phase diagram for $g_{2}^{2}=1.0$ in Fig.~\ref{g1}.
Since we have already shown that the physical phase gets rather narrower in
the strong-coupling limit in 4d lattice QCD in Sec.~\ref{sec:SLQ}.
it is natural to consider that the single-coupling GN model in Fig.~\ref{fig1} 
is sufficient to mimic 4d QCD at least for investigating the chiral phase diagram.
We note that we can introduce two more types of four-point interactions 
in the GN action (\ref{SGNS}) 
as $(\bar{\psi}_{n}i\gamma_{1}\psi_{n})^{2}$ and
$(\bar{\psi}_{n}i\gamma_{3}\psi_{n})^{2}$,
which give $\pi_{1}$ and $\pi$ meson fields.
Both vacuum expectation values of $\pi_{1}$ and $\pi$,
if they are nonzero, cause spontaneous parity symmetry breaking. 
However, by comparing the effective potentials or 
solving four gap equations
(${\delta \tilde{S}_{\rm eff}\over{\delta \sigma}}=0$,
${\delta \tilde{S}_{\rm eff}\over{\delta \pi_{1}}}=0$,
${\delta \tilde{S}_{\rm eff}\over{\delta \pi_{2}}}=0$, 
${\delta \tilde{S}_{\rm eff}\over{\delta \pi}}=0$),
we can clearly show that they have zero vacuum expectation values as 
in the case of the strong-coupling lattice QCD in Appendix.A.
Instead of solving all the gap equations numerically, 
we here show the consistency check on $\pi_{1}=0$. 
From two saddle-point equations ${\delta \tilde{S}_{\rm eff}\over{\delta \sigma}}=0$ 
and ${\delta \tilde{S}_{\rm eff}\over{\delta \pi_{1}}}=0$ we derive one equation,
\begin{equation}
0=2\int{d^{2}k\over{(2\pi)^{2}}}
{\sin k_{1}\over{\sigma^{2}+\pi^{2}+(\pi_{2}+(1+d)\sin k_{2}-\cos k_{1})^{2}+(\pi_{1}+\sin k_{1})^{2}}}.
\end{equation}
If we take $\pi_{1}=0$, the right-hand side becomes an odd function of $k_{1}$, 
and the above equation holds identically.
We can also show from the gap equations that $\pi=0$ is preferred.
These results mean that the parity-broken vacuum is not preferred, and 
as long as we consider the vacuum of the theory, 
the two four-fermi interactions in (\ref{SGNS}) are sufficient.
We next show that, however, we have to take into account the excitation of 
$\pi_{1}$ and $\pi$ to investigate the Lorentz symmetry restoration in the GN model.  

\begin{figure}
\includegraphics[height=7cm]{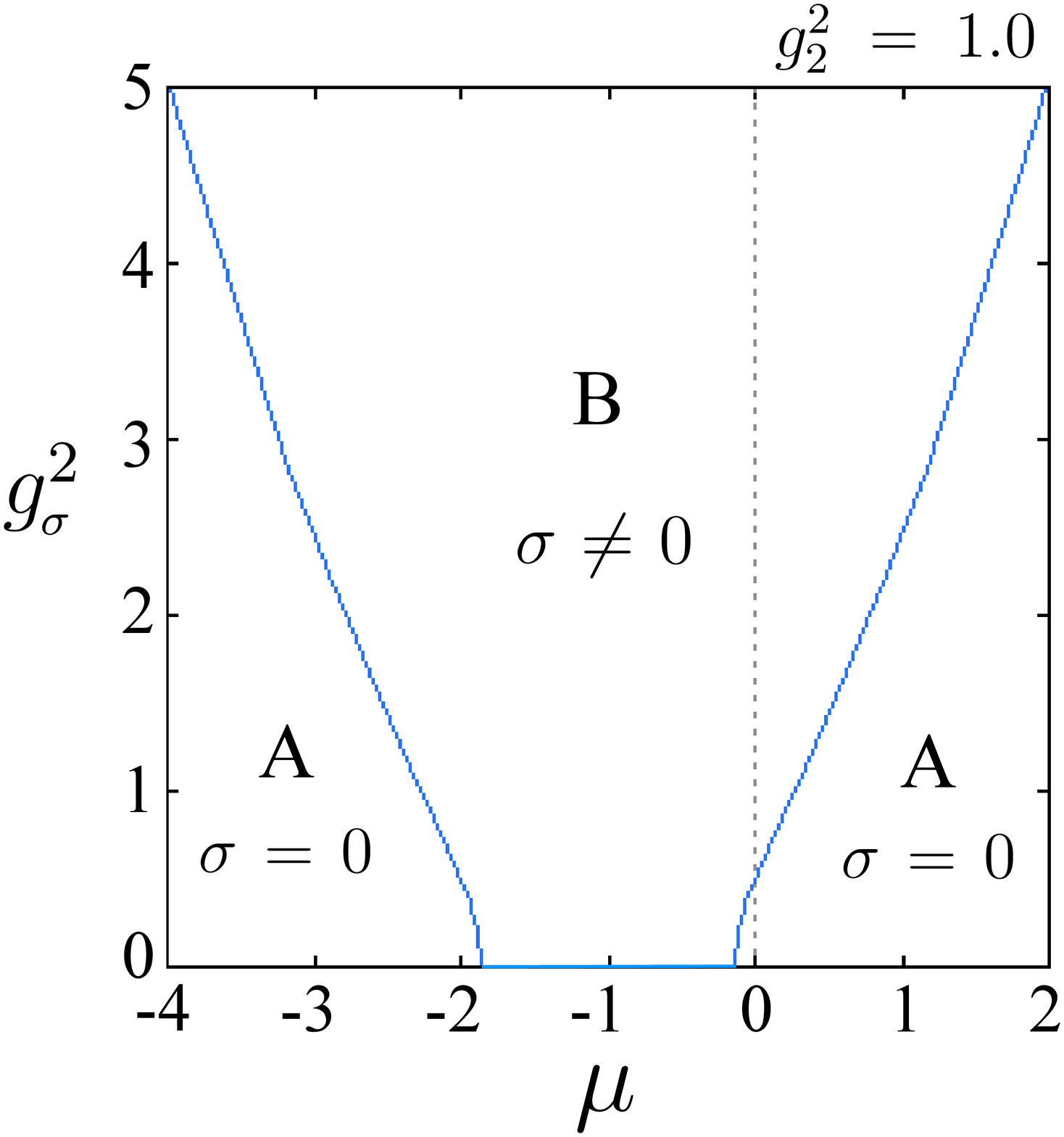} 
\caption{Chiral phase diagram for KW Gross-Neveu model in the $\mu$-$g_{\sigma}^{2}$ space
with $g_{2}^{2}=1.0$ ($d=0$). The physical (minimal-doubling) phase with chiral condensate gets larger toward the strong-coupling limit.}
\label{g1}
\end{figure}

By now we have considered only the vacuum of the theory. 
Although it works to elucidate the chiral phase structure, 
we need look into excitations from the vacuum in order to 
investigate renormalization for the rotation symmetry restoration.
By expanding the mesonic action up to the 2nd order,
we can derive the meson propagator matrix for $\sigma, \pi_{1}, \pi_{2}, \pi$
as in the strong-coupling lattice QCD in Sec.~\ref{sec:SLQ}.
By diagonalizing this $4\times 4$ matrix, 
we obtain the proper propagators.
Then, we can discuss how the rotation symmetry can be restored by
tuning the parameters including $\mu$, $d$, $g_{\sigma}^{2}$ and $g_{2}^{2}$.
Although we can perform this procedure in principle,
all the matrix components include complicated momentum integrals in this case.
Moreover, we need to substitute the VEV of meson fields derived from the
gap equations, which also require numerical integrals.
We consider that these numerical calculations are beyond the scope 
of this study, but the future work should be devoted to it.
Instead, we here show a process of deriving the dispersion relations
in details in the following. 
The meson excitations from the vacuum are given by
\begin{equation}
\sigma=\bar{\sigma}+\sigma(n),\,\,\,\,
\pi_{1}= \bar{\pi}_{1}+\pi_{1}(n),\,\,\,\,
\pi_{2}= \bar{\pi}_{2}+\pi_{2}(n),\,\,\,\,
\pi=\bar{\pi}+\pi(n),\,\,\,\,   
\end{equation}
where VEVs ($\bar\sigma$, $\bar\pi_{1}$, $\bar\pi_{2}$, $\bar\pi$) are 
determined by the gap equations.
We have already shown $\bar{\pi}_{1}=\bar{\pi}=0$ while $\bar{\sigma}$ and $\bar{\pi}_{2}$
depends on the parameters.
The Dirac operator with these excitations are written as
\begin{align}
D_{n,m}&=(\mathcal{S}_{0}^{-1})_{n,m}+\mathcal{M}(n)\delta_{n,m},
\\
(\mathcal{S}_{0}^{-1})_{n,m}&=[\bar{\sigma}+i\gamma_{2}\bar{\pi}_{2}]\delta_{n.m}+
{\gamma_{\mu}\over{2}}(\delta_{n+\mu,m}-\delta_{n-\mu,m})/a
\nonumber\\
&\,\,\,\,\,\,\,\,\,\,\,\,\,\,\,\,\,\,\,\,-{i\gamma_{2}\over{2}}(\delta_{n+\hat{1},m}+\delta_{n-\hat{1},m})/a
+d{\gamma_{2}\over{2}}(\delta_{n+\hat{2},m}-\delta_{n-\hat{2},m})/a
\\
\mathcal{M}(n)&=\sigma(n)+i\gamma_{1}\pi_{1}(n)+i\gamma_{2}\pi_{2}(n)+i\gamma_{3}\pi(n),
\end{align}
where we make the lattice spacing $a$ manifest.
Note $\sigma, \pi_{1},\pi_{2}, \pi \sim O(1/a)$.
For completeness, we consider the following action including four
independent coupling constants,
\begin{equation}
S_{\rm eff}=\sum_{n}[{1\over{2g_{\sigma}^{2}}}(\sigma-m)^{2}
+{1\over{2g_{1}^{2}}}\pi_{1}^{2}
+{1\over{2g_{2}^{2}}}(\pi_{2}-\mu-1)^{2}
+{1\over{2g_{\pi}^{2}}}\pi^{2}]-{\rm Tr}\,\log [\mathcal{S}_{0}^{-1}(1+\mathcal{S}_{0}\mathcal{M})]_{n,m}.
\end{equation}
$\rm Tr$ is a trace for coordinate and spinor spaces.
By expanding this expression up to second order of mesonic fluctuations, 
we derive the effective action up to the quadratic order as
\begin{equation}
S_{\rm eff}=S^{(0)}_{\rm eff}+S^{(2)}_{\rm eff}+\cdot\cdot\cdot,
\end{equation}
with
\begin{equation}
S^{(2)}_{\rm eff}=\sum_{n}\left[{\sigma(n)^{2}\over{2g_{\sigma}^{2}}}
+{\pi_{1}(n)^{2}\over{2g_{1}^{2}}}
+{\pi_{2}(n)^{2}\over{2g_{2}^{2}}}
+{\pi^{2}(n)\over{2g_{\pi}^{2}}}
+{1\over{2}}{\rm tr}\sum_{m}
\mathcal{S}_{0}(m,n)\mathcal{M}(n)\mathcal{S}_{0}(n,m)\mathcal{M}(m)\right],
\end{equation}
with $\rm{tr}$ is a trace for the spinor space.
By fourier transformation we can derive the form in the momentum space as
\begin{align}
S^{(2)}_{\rm eff}=\int_{-\pi/a}^{\pi/a} {d^{2}p\over{(2\pi)^{2}}} 
\Big[  {\sigma(p)^{2}\over{2g_{\sigma}^{2}}}+{\pi_{1}(p)^{2}\over{2g_{1}^{2}}} 
&+{\pi_{2}(p)^{2}\over{2g_{2}^{2}}} +{\pi(p)^{2}\over{2g_{\pi}^{2}}} 
\nonumber\\
&+ {1\over{2}}{\rm tr}\int_{\pi/a}^{\pi/a} {d^{2}k\over{(2\pi)^{2}}} \mathcal{S}_{0}(p+k)\mathcal{M}(p)
\mathcal{S}_{0}(k)\mathcal{M}(-p) \Big],
\end{align}
with
\begin{align}
&\mathcal{S}_{0}(k)={1\over{\det(\mathcal{S}_{0}^{-1}(k))}} \left[\bar{\sigma}-i\gamma_{2}\left(\bar{\pi}_{2}+(1+d){\sin ak_{2}\over{a}}+{\cos ak_{1}\over{a}}\right)-i\gamma_{1}{\sin ak_{1}\over{a}}\right],
\\
&\det(\mathcal{S}_{0}^{-1}(k)) =\bar{\sigma}^{2}
+\left(\bar{\pi}_{2}
+(1+d){\sin ak_{2}\over{a}}-{\cos ak_{1}\over{a}}\right)^{2}+
\left({\sin ak_{1}\over{a}}\right)^{2}.
\end{align}
Finally, the mesonic propagator matrix is given by
\begin{equation}
\mathcal{D}_{XY}(p)={\delta^{2} S_{\rm eff}^{(2)} 
\over{\delta \mathcal{M}_{X}(p)\delta \mathcal{M}_{Y}(-p)}}
\end{equation}
where $X$ and $Y$ stand for one of channels $\sigma$, $\pi_{1}$, $\pi_{2}$ and $\pi$. 
For example, $\mathcal{D}_{\sigma\sigma}(p)$ is given by
\begin{align}
\mathcal{D}_{\sigma\sigma}(p) =& {1\over{g_{\sigma}^{2}}}
+\int_{\pi/a}^{\pi/a} {d^{2}k\over{(2\pi)^{2}}}
{1\over{\det[\mathcal{S}_{0}^{-1}(p+k)\mathcal{S}_{0}^{-1}(k)]}} 
\times\Big[\bar{\sigma}^2
\nonumber\\
&-\Big(\bar{\pi}_{2}+(1+d){\sin ak_{2}\over{a}}+{\cos ak_{1}\over{a}}\Big)
\Big(\bar{\pi}_{2}+(1+d){\sin a(p_{2}+k_{2})\over{a}}+{\cos a(p_{1}+k_{1})\over{a}}\Big)
\nonumber\\
&-{\sin ak_{1}\sin a(p_{1}+k_{1})\over{a^{2}}}\Big].
\label{dead}
\end{align}
In the same way we can derive all other 16 components of the matrix, 
all of which take nonzero values in general.
What we are interested in is terms of the components up to $O(p^{2})$
as far as we consider the rotation symmetry up to $O(a)$ discretization errors.
The coefficients can be extracted through numerical integrals of
the equations as (\ref{dead}).
We note that we also need to substitute into the integral 
the values of VEV of $\sigma$ and $\pi_{2}$ derived from the gap equations.
Then we diagonalize the $4\times 4$ meson matrix $\mathcal{D}_{XY}$
to derive the proper meson propagator. 
By introducing $p=({\bf p},iE)$ into the propagators,
we find the dispersion relations.
The question is which parameters among $\mu$, $d$, $g_{\sigma}^{2}$, $g_{1}^{2}$, 
$g_{2}^{2}$, $g_{\pi}^{2}$ need to be tuned to recover Lorentz symmetry. 
It can give an important suggestion to lattice QCD with the minimal-doubling fermions.
We devote a future work to this analysis.


\section{Conjecture on phase structure in QCD}
\label{sec:QCD}

\begin{figure}
\includegraphics[height=7cm]{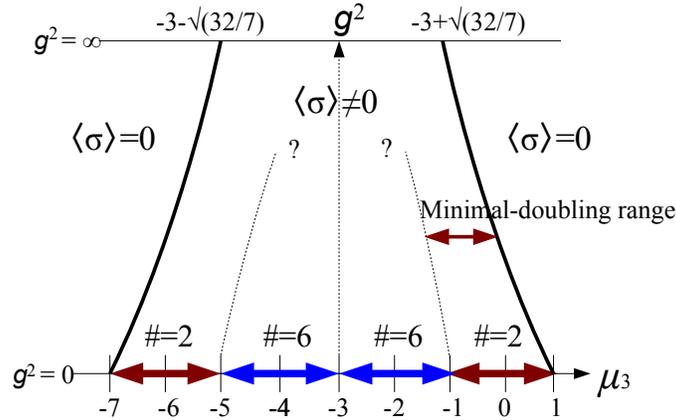} 
\caption{Conjecture on $\mu_{3}$-$g^{2}$ chiral phase structure 
for Karsten-Wilczek fermion with r=1.
The width of the minimal-doubling range determines
how hard it is to tune $\mu_{3}$.}
\label{AL}
\end{figure}

From the study of strong-coupling lattice QCD and the Gross-Neveu model
we speculate on the whole chiral phase structure in lattice QCD with 
Karsten-Wilczek fermion. 
Fig.~\ref{AL} is a conjectured chiral phase structure with the number of flavors 
in the $\mu_{3}$-$g^{2}$ space for $r=1$.
There are roughly two phases with and without chiral condensate, 
or equivalently with and without SSB of chiral symmetry.
As was shown in the previous section, the boundary between chiral symmetric 
and broken phases starts from the edge of the two-flavor region of the free theory. 
We expect that the chiral boundaries are connected to the 
two-flavor and no-flavor phases also in 4d QCD as shown in Fig.~\ref{AL}.

The question is a boundary between two-flavor and six-flavor ranges.
In the weak-coupling limit ($g^{2}=0$) we analytically know
the number of physical flavors: There are four sectors with two, six, six and two flavors. 
There are only no flavors of fermions outside these ranges. 
Toward the strong coupling, these ranges will change with $g^{2}$ as shown in Fig.~\ref{AL}. 
We have seen that we cannot distinguish 
two-flavor and six flavor ranges in the strong-coupling limit, 
which means that the number of species becomes an ambiguous 
notion in this limit. We thus expect that the boundary disappears 
at a certain gauge coupling, and the two-flavor and six-flavor regions become
undistinguishable as shown in Fig.~\ref{AL}.

From the viewpoints of practical application to two-flavor QCD,
the relevant parameter $\mu_{3}$ has to be tuned
to cancel the $O(1/a)$ imaginary chemical potential 
renormalization for the two flavors.
One necessary condition (but not a sufficient condition) for this purpose is to
set $\mu_{3}$ in the minimal-doubling range to realize the two-flavor QCD.
As we conjectured in Fig.~\ref{AL}, the minimal-doubling range 
in the middle gauge coupling should have some width.
One possible indicator of the minimal-doubling phase is the pion spectrum:
If $\mu_{3}$ is in the no-flavor range, there is no SSB of chiral symmetry
and no massless pion. If $\mu_{3}$ gets into the six-flavor region, 
the number of pseudo Nambu-Goldstone bosons increases.
However, setting $\mu_{3}$ in the minimal-doubling range 
is not sufficient for physical QCD to be described:
The Lorentz symmetric dispersion relation is broken down 
for general values of $\mu_{3}$ in the minimal-coupling range 
as shown in Eq.~(\ref{dp1}) and below for a free theory. 
\begin{equation}
D(p)\sim i\gamma_{i}p_{i}+ i\gamma_{4}p_{4} \sqrt{(1+d_{4})^{2}-\mu_{3}^{2}} \,+\,O(ap^{2}).
\label{dp2}
\end{equation}
Since this free-theory argument indicates that the rotation symmetry can be 
restored by tuning $d_{4}$ as $(1+d_{4})^{2}-\mu_{3}^{2}=1$,
we may be able to restore Lorentz symmetry just by
tuning $d_{4}$ with $\mu_{3}$ being set in the minimal-doubling range.
Note that the minimal-doubling range gets larger with nonzero $d_{4}$
as shown in Sec.III and IV, thus it seems that proper tuning of $d_{4}$ can be done
for any value of $\mu_{3}$ without breaking down minimal-doubling.
(We need one more parameter tuning for the plaquette action in any case.)
We also emphasize that the same relative tuning of $\mu_{3}$ and $d_{4}$ makes 
tree level couplings of the gauge field to the fermions have a correct 
Lorentz-symmetric form. To show this, we look into
the quark-quark-gluon vertex at the tree level.
For the case of $\mu_{3}=0$ and $d_{4}=0$ it is given by
\begin{equation}
V(p,k)=-ig_{0}\left( \gamma_{\mu}\cos {a(p_{\mu}+k_{\mu})\over{2}}
+\gamma_{4}(1-\delta_{\mu4})\sin {a(p_{\mu}+k_{\mu})\over{2}} \right),
\end{equation}
as shown in \cite{Cap3}.
For nonzero $\mu_{3}$ and $d_{4}$, it is modified as
\begin{equation}
V(p,k)=-ig_{0}\left( \sum_{j=1}^{3}\gamma_{j}\cos {a(p_{j}+k_{j})\over{2}}
+ \gamma_{4}\left[(1+d_{4})\cos {a(p_{4}+k_{4})\over{2}}
+\sum_{i=1}^{3}\sin{a(p_{i}+k_{i})\over{2}}\right] \right),
\label{mdV}
\end{equation} 
where we have no direct emergence of $\mu_{3}$ since it is a parameter 
for onsite (non-hopping) terms.
However, as we discussed, the zeros of the Dirac operator for nonzero $\mu_{3}$ and $d_{4}$ 
is given by a function of $\mu_{3}$ and $d_{4}$ 
as $\bar{p}=\bar{k}=(0, 0, 0, {1\over{a}}\arcsin(-{\mu_{3}\over{1+d_{4}}}))$.
Now we expand both $p$ and $k$ about the zeros as $p\to \bar{p}+p$ and $k\to \bar{k}+k$. 
In particular, the coefficient of $\gamma_{4}$ in (\ref{mdV}) is expanded as
\begin{align}
&(1+d_{4})\Big[\cos {a(\bar{p}_{4}+\bar{k}_{4})\over{2}} \cos {a(p_{4}+k_{4})\over{2}}
-\sin {a(\bar{p}_{4}+\bar{k}_{4})\over{2}} \sin {a(p_{4}+k_{4})\over{2}}\Big]
+\sum_{i=1}^{3}\sin {a(p_{i}+k_{i})\over{2}} 
\nonumber\\
=&(1+d_{4})\Big[\sqrt{1-{\mu_{3}\over{(1+d_{4})^{2}}}} \cos {a(p_{4}+k_{4})\over{2}}
+ {\mu_{3}\over{1+d_{4}}} \sin {a(p_{4}+k_{4})\over{2}}\Big]
+\sum_{i=1}^{3}\sin {a(p_{i}+k_{i})\over{2}} 
\nonumber\\
=&(1+d_{4})\sqrt{1-{\mu_{3}\over{(1+d_{4})^{2}}}}+O(ap, ak),
\end{align}
Then, the vertex surviving in the naive continuum limit is given by
\begin{equation}
V(p,k)=-ig_{0}\left(\gamma_{1}+\gamma_{2}+\gamma_{3}
+\gamma_{4}\sqrt{(1+d_{4})^2-\mu_{3}^{2}} \right)+O(ap,ak).
\end{equation} 
Here we omit the $\pm$ sign in front of $\gamma_{4}$
for the doubler pairs for simplicity.
It is now obvious that the tuning condition for the speed of light 
$(1+d_{4})^{2}-\mu_{3}^{2}=1$ also fixes the couplings of the gauge fields
to the fermion fields in the tree level up to the discretization errors.
At least in the naive continuum limit, we can have a correct set of
the Feynman rules for fermion fields with the condition.
However it is too early to conclude that this condition is sufficient
for Lorentz symmetry restoration since all the other Ward identities may not 
be corrected by it in the interacting theory including the loop effects:
To discuss details, we consider the quark self-energy in lattice QCD
with minimal-doubling fermions following \cite{Cap3} as
\begin{equation}
\Sigma(p,m)=i\gamma_{\mu}p_{\mu} \Sigma_{1}(p)+m\Sigma_{2}(p)
+d_{1}(g_{0})\cdot i\gamma_{4}p_{4}+d_{2}(g_{0})\cdot i{\gamma_{4}\over{a}},
\end{equation} 
where $\Sigma_{1}$, $\Sigma_{2}$, $d_{1}$ and $d_{2}$ can be calculated 
in the perturbative analysis.
It is clear that $\mu_{3}$ and $d_{4}$ corresponds to counter parameters for
$d_{2}$ and $d_{1}$ respectively, and the Dirac operator (\ref{dp2}) is renormalized
as $d_{4}\to d_{4}+d_{1}$ and $\mu_{3}\to \mu_{3}+d_{2}$ in the interacting theory.
The last term with $d_{2}$, or the $O(1/a)$ renormalization, causes 
a shift of the poles of the Dirac propagator away from their original positions
as well as the change of the speed of light as shown in (\ref{dp2}).
In the present work, we have also shown that this contribution changes
the size of the minimal-doubling range at the finite gauge coupling 
and provides the non-trivial phase structure as shown in Fig.~\ref{AL}.
The question is whether or not we need to move the poles of the propagator
back to the tree-level positions by tuning $\mu_{3}$ for a correct continuum limit.
As far as the dispersion relation can be restored by $d_{4}$, 
it seems that the position of poles is not relevant to physics.
However, in practical use of the minimal-doubling fermion, $d_{4}$ should be also
tuned to make the conserved charge unity as shown in \cite{Cap3}.
It is not obvious whether this condition can also fix the speed of light non-perturbatively.
More generally speaking, it is very nontrivial whether all the Ward identities
are fixed only by one tuning condition in the interacting theory. 
If we cannot restore Lorentz symmetry only with $d_{4}$ tuning unlike the free theory, 
it means that we still need to fine-tune the three parameters independently 
for a correct continuum limit of lattice QCD simulations 
\cite{Cap1, Cap2, Cap3}.
Further study is needed to figure out this point.

In the end of this section, we comment on another possibility for studying
minimal-doubling fermions.
One interesting possibility is the chiral perturbation theory for
minimal-doubling fermions.
Although we expect that it is quite tedious to construct the minimal-doubling 
ChPT with the lower discrete symmetry than Wilson and staggered fermions, 
the process could have some similarities with that of the in-medium ChPT \cite{L}.
If we succeed to construct the minimal-doubling ChPT, it is intriguing
to consider the Lorentz symmetry restoration within the theory and discuss
the parameter tuning for the symmetry restoration.
We can also investigate the vacuum and the phase structure in the theory.
We devote future works to the study on the minimal-doubling ChPT.


\section{Summary and Discussion}
\label{sec:SD}

In this paper we investigate the chiral phase structure in the parameter space 
for lattice QCD with minimal-doubling fermions, which can be seen as
lattice fermions with a species-dependent naive chemical potential term.
We study the phase structure with Karsten-Wilczek fermion 
by using strong-coupling lattice QCD and the Gross-Neveu model, 
and find out the nontrivial chiral phase structure in the $\mu_{3}-g^{2}$ plane.

In Sec.~\ref{sec:FCP}, we have proposed flavored-chemical-potential
lattice fermions, where some of doublers are eliminated by a species-dependent
chemical potential term without losing all chiral symmetries.
Minimal-doubling fermions are shown to be a special case of this type.
In Sec.~\ref{sec:SLQ} we investigate the chiral phase structure of lattice QCD with
Karsten-Wilczek fermions in the strong-coupling regime.
We derive an effective action for the scalar and 4th vector fields, and 
find that chiral symmetry is spontaneously broken in a certain range of the 
relevant parameter $\mu_{3}$
while the chiral condensate becomes zero outside the range.
We show that there is a 2nd-order phase transition between 
chiral symmetric and broken phases as a function of $\mu_{3}$.
We also show that pion becomes massless as a Nambu-Goldstone boson 
in the chiral-broken phase while the scalar meson becomes massless only 
on the second-order phase boundary due to the critical behavior. 
In Sec.~\ref{sec:GN} we obtain information on the whole chiral phase structure in 
the $\mu_{3}-g^{2}$ space by using the Gross-Neveu model with KW fermion
in large N limit. From the gap equations we derive a chiral phase diagram.
In Sec~\ref{sec:QCD} we discuss the whole phase structure in 4-dimensional
lattice QCD with KW fermions. We conjecture the chiral phase structure
and numbers of massless flavors in the phase diagram.  
We also discuss the fine-tuning process of $\mu_{3}$ in lattice QCD 
from the viewpoint of the minimal-doubling range 
of the conjectured phase diagram.

In this paper we have investigated whether or not chiral symmetry is 
spontaneously broken depending on the parameters.
Unlike Wilson parity-flavor breaking phase with the width $m\sim O(a^3)$ 
\cite{SS}, it seems that we do not have fine phase structures in the 
weak-coupling regime as shown in Fig.~\ref{AL}.
It is because the chiral symmetry breaking is physical, and
there is no competition between the relevant parameter and
the lattice artifact, which did cause the Aoki phase or
the Creutz-Sharpe-Singleton phase in Wilson fermion \cite{SS,Creutz3}.   
Although our analysis on chiral symmetry and that on parity
symmetry in Wilson have some common features,
they are on the different levels.

In this work, we do not take much care of $\pi_{4}$ condensate
nor its physical implication. As we discussed in Sec.~\ref{sec:SLQ}
this condensate is likely to be related to the (pseudo) quark density.
On the other hand, Eq.~(\ref{g1}) in the strong-coupling limit shows that $\pi_{4}$ is nonzero for any 
values of $\mu_{3}$ and $d_{4}$ except for the unphysical point $\mu_{3}+3r=0$.
We consider that it is just a strong-coupling artifact, and the $\pi_{4}$ 
condensate can be eliminated by fine-tuning all the
three parameters appropriately in the weak coupling.
There is possibility that another way of introducing the dimension-4 counterterm 
as an O(1) imaginary chemical potential term
$\bar{\psi}_{n}\gamma_{4}(e^{i\mu}\psi_{n+4}-e^{-i\mu}\psi_{n-4})$
may work since the flavored-chemical-potential 
term could generate $O(1)$ effective imaginary chemical potential too.

In the end of this paper,
we refer to a future work.
We can apply the flavored-chemical-potential (FCP) fermions to finite-temperature 
and finite-density lattice QCD, and obtain finite-($T$,$\mu$) QCD phase diagram.
In our next work \cite{MKO}, we will propose a new method to
study lattice QCD in medium by using FCP fermion formulations.


\begin{acknowledgments}
TM is thankful to M.~Creutz, T.~Kimura and A.~Ohnishi 
for fruitful discussion and hearty encouragement.
The author also thanks F.~Karsch for the discussion.
The author appreciates the technical help by T.~Kawanai. 
TM is supported by Grant-in-Aid for Japan Society for the Promotion of Science (JSPS) 
Postdoctoral Fellows for Research Abroad(No.24-8).
\end{acknowledgments}

\appendix

\section{Possibility of pion condensation}
We here assume a form of meson condensate with
chiral, 4th vector and pion condensates as
\begin{equation}
\mathcal{M}_0 =\sigma{\bf 1}_4 + i \pi_4\gamma_4 + i\pi_{5}\gamma_{5}.
\end{equation} 
The effective potential for $\sigma$, $\pi_{4}$ and $\pi_{5}$ is
given by
\begin{align}
\mathcal{V}_{\rm eff}(\sigma,\pi_{4}, \pi_{5})&=
{1\over{2}}\log (\sigma^{2}+\pi_{4}^{2}+\pi_{5}^{2})-m\sigma +(\mu_{3}+3r)\pi_{4}
\nonumber\\
&-{1\over{4}}[3(1+r^2)+(1+d_{4})^{2}]\sigma^{2}
-{1\over{4}}[3(1-r^2)-(1+d_{4})^{2}]\pi_{4}^{2} - {1\over{4}}[3(1+r^2)+(1+d_{4})^{2}]\pi_{5}^{2}.
\label{V5}
\end{align}
We now find saddle points of $\mathcal{V}_{\rm eff}$ given by
\begin{eqnarray}
\frac{3(1+r^2)+(1+d_{4})^{2}}{2}\sigma + m -\frac{\sigma}{\sigma^2+\pi_4^2 + \pi_{5}^{2}} &=& 0 \, ,
\\
\frac{3(1-r^2)-(1+d_{4})^{2}}{2}\pi_4 -(\mu_3+3r) -\frac{\pi_4}{\sigma^2+\pi_4^2 +\pi_{5}^{2}} &=& 0 \,,
\label{p4eq}
\\
\frac{3(1+r^2)+(1+d_{4})^{2}}{2}\pi_{5} -\frac{\pi_{5}}{\sigma^2+\pi_4^2 +\pi_{5}^{2}} &=& 0.
\label{pi5eq}
\end{eqnarray}
We now consider a case for $r=1$, $d_{4}=0$ and $m=0$.
We have two types of solutions as $\pi_{5}=0$ and $\pi_{5}\not=0$.
For $\pi_{5}=0$, we have the two solutions $\mathcal{M}_{0}^{A}$ and $\mathcal{M}_{0}^{B}$
in (\ref{chis})(\ref{chib}) as we discussed in Sec.~\ref{sec:SLQ}.
For $\pi_{5}\not=0$, (\ref{p4eq})(\ref{pi5eq}) give $\sigma^2+\pi_4^2 +\pi_{5}^{2}=2/7$ 
and $\pi_{4}=-\bar{\mu}_{3}/4=-(\mu_{3}+3)/4$.
By substituting them into (\ref{V5})
the effective potential as a function of $\bar{\mu}_{3}$ is given by
\begin{equation}
\mathcal{V}_{\rm eff}(\pi_{5}\not=0) 
= {1\over{2}}\log {2\over{7}} -{1\over{2}}-{1\over{8}}\bar{\mu}_{3}^{2}.
\end{equation}
On the other hand, for example, the effective potential for one of $\pi_{5}=0$ solutions 
$\mathcal{M}_{0}^{B}$ is given by
\begin{equation}
\mathcal{V}_{\rm eff}(\pi_{5}=0, \mathcal{M}_{0}^{B}) 
= {1\over{2}}\log {2\over{7}} -{1\over{2}}-{1\over{8}}\bar{\mu}_{3}^{2}.
\end{equation}
It is obvious $\mathcal{V}_{\rm eff}(\pi_{5}\not=0) \not< \mathcal{V}_{\rm eff}(\pi_{5}=0, \mathcal{M}_{0}^{B})$ for any value of $\bar{\mu}_{3}$.
It indicates that the $\pi_{5}\not=0$ solution is unlikely to be a vacuum.
We thus conclude that, at least in this framework, there is no pion 
condensate or no spontaneous parity symmetry breaking.
We also perform the same analysis for the condensate form
$\sigma{\bf 1}_4 + i \pi_4\gamma_4 + i\pi_{45}\gamma_{45}$,
and will find that $\pi_{45}\not=0$ is not a vacuum of the theory.

\section{Creutz-Borici case}
A free action of Creutz-Borici fermion is given by
\begin{align}
S_\mrm{BC}&=\sum_{n}[{1\over{2}}\sum_{\mu}\bar{\psi}_{n}(\psi_{n+\mu}-\psi_{n-\mu})+
\frac{i  r}{2}\sum_{\mu}\bar{\psi}_n(\Gamma-\gamma_\mu)\left(2\psi_n-\psi_{n+\hat{\mu}}-\psi_{n-\hat{\mu}}\right)  
\nonumber\\
&\,\,\,\,\,\,\,\,\,\,\,\,\,\,\,\,\,+ i  c_3\bar{\psi}_n\Gamma\psi_n + m\bar{\psi}_n\psi_n ] \,,\\
\Gamma &=\frac{1}{2}\sum_\mu \gamma_\mu, \quad \gamma_\mu^\prime =\Gamma\gamma_\mu\Gamma =\Gamma-\gamma_\mu,
\end{align}
where $c_{3}$ corresponds to $\mu_{3}$ in Karsten-Wilczek fermion and
$\Gamma$ satisfies $\Gamma^2=1$ and $\{ \Gamma,\gamma_\mu\} = 1$.
In this case we have onsite operator and projection operators as 
$\hat M = m{\bf 1}_4 + i (c_3-2r)\Gamma^\T$ and
\begin{eqnarray}
P^+_\mu &=& \frac{1}{2} \{\gamma_\mu (1+i  r) +i r\Gamma \}, \quad
P^-_\mu = \frac{1}{2} \{\gamma_\mu (1-i  r) -  ir\Gamma \} .
\end{eqnarray}
The strong-coupling analysis is done in a parallel manner to KW fermion.
By taking $\mathcal{M} = \sigma + i \pi_\Gamma \Gamma$, 
the corresponding gap equations become
\begin{eqnarray}
 2(1+r^2) \sigma +m - \frac{\sigma}{\sigma^2+\pi_\Gamma^2} &=& 0 \, ,\\
 (1+r^2)\pi_\Gamma -(c_3+2 r) - \frac{\pi_\Gamma}{\sigma^2+\pi_\Gamma^2} &=& 0 \, , 
\end{eqnarray}
For $m=0$, equations for second-order phase boundaries are given by
\begin{eqnarray}
 2(1+r^2)  - \frac{1}{\pi_\Gamma^2} &=& 0 \, ,\\
 (1+r^2)\pi_\Gamma -(c_3+2 r) - \frac{1}{\pi_\Gamma} &=& 0 \, , 
\end{eqnarray}   
Then chiral boundaries are given by
\begin{equation}
c_{3}=\pm{\sqrt{1+r^{2}}\over{\sqrt{2}}}-2r.
\end{equation}
For $r=1$ chiral condensate is nonzero and chiral symmetry is spontaneously
broken for $-3<c_{3}<-1$.


\end{document}